\documentclass[
reprint,
showpacs,
amsmath,amssymb,
aps,
prb,
floatfix,
]{revtex4-1}

\usepackage{graphicx}
\usepackage[colorlinks=true,citecolor=blue,linkcolor=magenta]{hyperref}

\newcommand{\kg}{{kagome }}
\newcommand{\st}{sawtooth }

\newcommand{\St}{Sawtooth }

\newcommand{\ssm}{\scriptscriptstyle\rm}
\renewcommand{\phi}{\varphi}
\renewcommand{\theta}{\vartheta}
\newcommand{\pdag}{\phantom{\dag}}

\def \be{\begin{equation}}
\def \ee{\end{equation}}
\def \ba{\begin{array}}
\def \ea{\end{array}}
\def \bea{\begin{eqnarray}}
\def \eea{\end{eqnarray}}
\def \nn{\nonumber}
\def \half{\frac{1}{2}}

\def \bG{{\bf G}}
\def \bQ{{\bf Q}}
\def \bK{{\bf K}}
\def \bR{{\bf R}}
\def \br{{\bf r}}

\def \bk{{\bf k}}

\def \bx{{\bf x}}

\def \a{{\alpha}}
\def \t{{\theta}}

\def \D{{\Delta}}
\def \d{{\delta}}

\def \f{{\varphi}}

\def \nd{{^{\vphantom{\dagger}}}}
\def \yd{^\dagger}
\def \av#1{{\langle#1\rangle}}
\def \ket#1{{\,|\,#1\,\rangle\,}}

\begin{document}
\title{Bose condensation in flat bands}

\author{Sebastian D. Huber}
\author{Ehud Altman}

\affiliation{Department of Condensed Matter Physics, The Weizmann
Institute of Science, Rehovot, 76100, Israel}

\date{\today}

\begin{abstract}
We derive effective Hamiltonians for lattice bosons with strong geometrical frustration of the kinetic energy by projecting the interactions on the flat lowest Bloch band. Specifically, we consider the Bose Hubbard model on the one-dimensional \st lattice and the two-dimensional \kg lattice. Starting from a strictly local interaction the projection gives rise to effective long-range terms stabilizing a supersolid phase at densities above $\nu_c=1/9$ of the \kg lattice. In the sawtooth lattice on the other hand we show that the solid order, which exists at the magic filling $\nu_c=1/4$, is unstable to further doping. The universal low-energy properties at filling $1/4+\d\nu$ are described by the well-known commensurate-incommensurate transition. We support the analytic results by detailed numerical calculations using the density-matrix renormalization group and exact diagonalization. Finally, we discuss possible realizations of the models using ultracold atoms as well as frustrated quantum magnets in high magnetic fields. We compute the momentum distribution and the noise correlations, that can be extracted from time of flight experiments or neutron scattering, and point to signatures of the unique supersolid phase of the \kg lattice.
\end{abstract}


\maketitle

\section{Introduction}

Strong geometric frustration can prevent straightforward ordering and thus lead to the emergence of novel highly correlated ground states. The best known examples of this phenomenon are from spin systems. Frustration of the magnetic exchange interactions on certain lattices gives rise to extensive degeneracy of classically ordered states,\cite{Moessner01b,Wannier50,Houtapple50,Syozi51,Villain80a,Chalker92,Ramirez94,Moessner98,Gardner10} invalidating a direct semiclassical spin-wave analysis. This picture has close analogy in the physics of the fractional quantum hall effect, where the huge degeneracy of a partially filled Landau level invalidates perturbative analysis in the interactions. In both cases the true ground state, which could be a Laughlin state, a spin liquid or some unexpected broken symmetry state, emerges from the degenerate manifold in a highly nontrivial way.

In this paper we address a related question concerning the ground states of weakly interacting bosons in a lattice which fully frustrates the bosons' kinetic energy. The usual expectation is that weakly interacting bosons will form a condensate in the lowest-energy single-particle state, or in other words, the lowest eigenstate of the kinetic-energy operator. However, if the hopping matrix elements on the lattice are sufficiently frustrated, the lowest Bloch band becomes flat, thus providing a huge degeneracy of single-particle states to which the bosons may condense.  The nature of the ground state is now fully determined by the interactions acting within the hugely degenerate manifold. Under these conditions a straightforward perturbative treatment in the interaction is of no use. The problem is inherently strongly correlated and provides an interesting route for understanding and perhaps even realizing novel phases of matter.

\begin{figure}[b]
\includegraphics{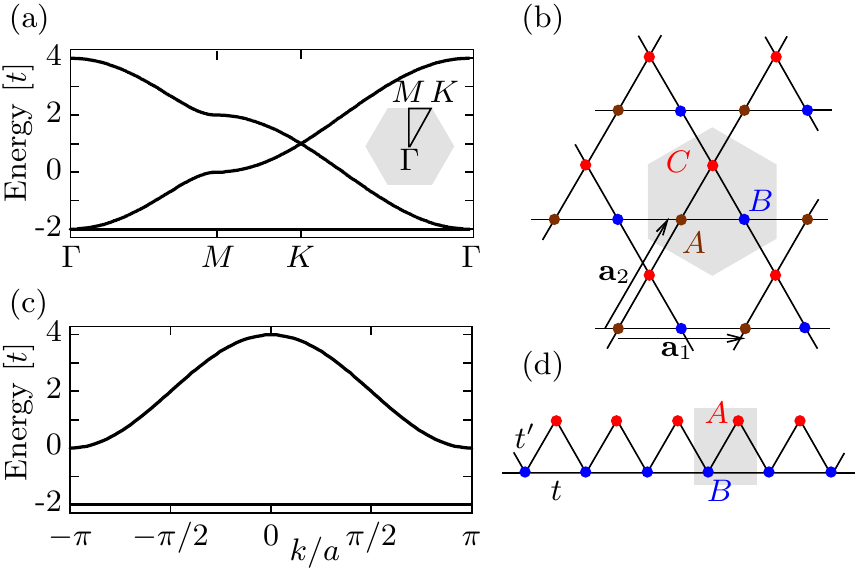}
\caption{
(color online)  (a) Single particle dispersion on the \kg lattice along high symmetry lines in the Brillouin zone (gray). (b) The \kg geometry with its lattice vectors ${\bf a}_{1/2}$ and the basis sites $A$, $B$, and $C$ in each unit cell (gray). (c) Single particle dispersion on the \st lattice as a function of momentum $k$ for $t'=\sqrt{2}t$. (d) The \st geometry with couplings $t$ and $t'$ and the basis sites $A$ and $B$ in the unit cell (gray).
}
\label{fig:models}
\end{figure}

We shall specifically consider a Hamiltonian of the form
\begin{equation}
\label{eqn:bosonic-hamiltonian}
H =\sum_{\langle ij\rangle} |t_{ij}|
\bigl[ b_{i}^{\dag}b_{j}^{\pdag} + {\rm H.c}\bigr]
+\frac{U}{2}\sum_{i}b_{i}^{\dag}b_{i}^{\dag}b_{i}^{\pdag}b_{i}^{\pdag},
\end{equation}
where $b_{i}$ are bosonic operators defined on the sites of the two-dimensional\kg lattice. This model gives a flat lower Bloch band in the single particle spectrum.\cite{Zhitomirsky04,Zhitomirsky05,Bergman08} We shall also consider a related one-dimensionalmodel defined on the sawtooth lattice. Both models and the band structure they give rise to are depicted in Fig.~\ref{fig:models}.

Such models of bosons with flat bands are of direct relevance to real physical systems. Recently a number of proposals were put forward for realization of models with frustrated hopping using ultracold atoms in optical lattices.\cite{Ruostekoski09,Damski05} Another natural realization involves\textbf{\textbf{\textbf{}}} frustrated spin-1 magnets.  If the Curie-Weiss temperature is sufficiently low, as in $m$-MPYNN$\cdot$BF$_{4}$ (Refs. \onlinecite{Awaga94} and \onlinecite{Wada97}) ($\Theta_{\ssm CW}\!\approx\!3\,{\rm K}$),  the spins can be fully polarized, or nearly so, by external magnetic fields. The dilute population of magnons, or depolarized spins, in the highly polarized regime is well described by Hamiltonian (\ref{eqn:bosonic-hamiltonian}).

\begin{figure}[t]
\includegraphics{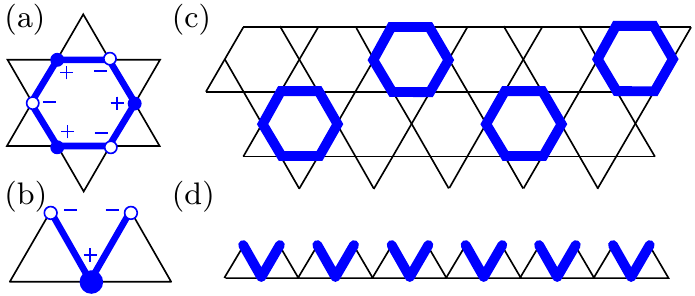}
\caption{
(color online) Strictly local eigenstates of the kinetic energy (local magnon states). (a) Resonating hexagon state on the \kg lattice where the $+$ and $-$ refer to the wave-function amplitude on each site around the hexagon. (b) Localized V-state on the sawtooth chain. (c) Ground state of the many-magnon system on the \kg lattice at density $\nu_{c}=1/9$. (d) Ground state of the many-magnon system on the \st chain at $\nu_{c}=1/4$.
}
\label{fig:strictly}
\end{figure}

The presence of a flat band implies the existence of localized eigenstates of the kinetic energy, as illustrated in  Fig.~\ref{fig:strictly}(a) and~\ref{fig:strictly}(b) for the \kg and sawtooth lattices. At sufficiently low filling of the lattice one can construct  exact many-body ground states of Hamiltonian (\ref{eqn:bosonic-hamiltonian}) from the localized single particle states.\cite{Schulenburg02,Zhitomirsky04,Derzhko09} This is done by occupying some subset of spatially nonoverlapping localized states. Of course the ground state in this case is massively degenerate, because of the many possible arrangements of nonoverlapping localized orbitals. However, there is a critical density $\nu_c$ at which the occupied localized states are closely packed, making an ordered crystal structure that breaks the lattice symmetry as shown in Fig.~\ref{fig:strictly}(c,d). The critical density is $\nu_{c}=1/9$ for the \kg and $\nu_{c}=1/4$ for the \st lattice, respectively. Formation of the close packed crystals is expected to give rise to magnetization plateaux in quantum magnets\cite{Hida01} and to incompressible insulators in ultracold atomic systems. Any additional bosons on top of the close packed configurations must overlap spatially with other bosons and the exact construction then no longer works.

The question we address in this paper is what ground states form at lattice filling slightly above $\nu_c$.  In particular, do the added atoms condense to form a superfluid on top of the underlying density wave? Such condensation is unusual because it would be entirely driven by the interactions rather than by the hopping. Another question is whether delocalization of the added atoms leads to destruction of the solid order.

Our general strategy to treat these systems is to project Hamiltonian (\ref{eqn:bosonic-hamiltonian}) onto the flat Bloch band and to obtain in this way an effective low-energy Hamiltonian which depends only on the (weak) interaction $U$. This approach is in the same spirit in which projection to the lowest Landau level is used to derive effective theories of quantum Hall states.\cite{Girvin84} The resulting low-energy model is defined on a new lattice, and is not frustrated. We are therefore able to analyze it using standard methods, such as bosonization and mean-field theories. The results are then checked against numerical calculations performed on the original Hamiltonian (\ref{eqn:bosonic-hamiltonian}) defined on the \kg and \st lattices.

Before proceeding we make note on previous work dealing with lattice boson models such as Eq. (\ref{eqn:bosonic-hamiltonian}) in the regime of large number of bosons per site $\nu\gg 1$, which is the opposite of the regime we consider here.\cite{Huse92,*Korshunov02} If $\nu\gg |t|/U > 1$, then Hamiltonian (\ref{eqn:bosonic-hamiltonian}) can be mapped to an effective Josephson array model. In the classical limit (where the charging energy can be neglected) the ground state manifold can be mapped onto a height model at its roughening transition. As a consequence, there is no long-range order in the phase $\langle e^{i\phi}\rangle$, but a ``tripled'' condensate forms where $\langle e^{3i\phi}\rangle$ exhibits long-range order.\cite{Huse92,*Korshunov02}

Below we lay out the structure of the paper and give a brief overview of our main results.  In Sec.~\ref{sec:sawtooth} we consider the \st model. The one-dimensional character of the chain allows to introduce the key ideas in a simple setup before applying them to the more complicated \kg lattice. The crucial step is to define an orthonormal basis of Wannier functions, which reside in the flat band.  Projection of the Hamiltonian on the flat band then takes the form of a spin-1/2 model with the spins located on the center of the Wannier functions and representing occupation of these states by bosons. Next, we take the long wave-length limit in a harmonic fluid approach.\cite{Haldane81} This analysis shows that upon increasing the density away from the magic filling $\nu_c=1/4$, the charge density wave is immediately destroyed due to proliferation of domain walls, which gain their mobility from the interaction term. The transition is in the universality class of the commensurate-incommensurate transition,\cite{Pokrovsky79,Schulz80,Chitra97} from which several sharp predictions follow. First, the delta-function peak in the density correlations at wave-vector $\pi/a$, which appears at $1/4$ filling, splits into a pair of power-law singularities at $(\pi\pm 2\delta\nu)/a$, where $\delta\nu=\nu-1/4>0$. Second, the established condensate is characterized by the Luttinger parameter $\tilde K=(1+\mathcal{D}\d\nu)/4$, which controls the power-law singularities. Here $\mathcal{D}$ is a positive constant independent of $\nu$. Note, that the value of $\tilde K=1/4$ at the transition point is quite remarkable for a system of bosons with purely on-site interactions and is a direct consequence of the geometric frustration. We find all of these predictions to be in good agreement with numerical density-matrix renormalization group (DMRG) calculations performed on the original \st model.

In Sec.~\ref{sec:kagome} we turn to the \kg lattice, focusing on filling $\nu\geq\nu_c=1/9$. Added bosons may either form mobile domain walls in the solid order or alternatively move as independent particles on interstitial sites of the density wave. Within an effective theory for these defects we estimate that the energy of a mobile interstitial, which does not destroy the density wave order, is lower.  To substantiate this result we again derive an effective low-energy model by projecting Hamiltonian (\ref{eqn:bosonic-hamiltonian}) onto the flat band. The resulting spin-$1/2$ model, written in terms of occupation of the flat band Wannier states, lives on the triangular lattice defined by the mid points of hexagons in the original \kg lattice. Because of the absence of a gap in the single particle spectrum [see Fig.~\ref{fig:models}(a)],
the justification of the low-energy theory is not as obvious in this case as in the \st model. However, we show that interaction with the uniform component of the density in the flat band generates a gap to excitation of the second band, which justifies the projection scheme.  A mean-field treatment of the low-energy Hamiltonian yields a ``lattice supersolid'' phase in the intermediate density regime $\nu_{c}<\nu<\nu_{\ssm MF}$, in which both the charge density wave (CDW) order and the off-diagonal order are present. The phase ordering structure is in the form of a 120$^{\circ}$ pattern on the $\sqrt{3}\!\times\!\sqrt{3}$ unit-cell of the CDW. In Sec.~\ref{sec:experiment} we discuss the experimental implications of our finding for both the cold-atoms and the spin-1 implementations.

\section{\St chain}
\label{sec:sawtooth}

We start with the discussion of the \st chain model\cite{Zhitomirsky04,Zhitomirsky05,Derzhko10} [cf. Fig.~\ref{fig:models}(d)], which can be viewed as a one-dimensional analogue of the \kg lattice. It is similarly constructed from corner-sharing triangles and can support a flat band for a specific ratio between the hopping matrix elements along the base and edges of the triangles.

\subsection{Flat band}
\label{sec:stflat}

Let us review the features of a tight-binding model on the \st lattice. We shall obtain the ratio of coupling constants required to give a flat lower band and define basis sets of localized states residing in this band. This will be important for implementing the projection on the flat band in the next section.

The hopping Hamiltonian can be formulated as a $2\times2$ matrix in terms of the ``spinor'' $\vec b_{k}=[b_{B,k},b_{A,k}]^{\rm T}$ representing the two sites in the unit cell:
\begin{equation}
\label{eqn:sthopping}
H_{\ssm kin}=\sum_{k}
\vec b_{k}^{\phantom{'}\dag}
\begin{bmatrix}
2t \cos(ka) & t'(1+e^{ika}) \\
t'(1+e^{-ika})& 0
\end{bmatrix}
\vec b_{k}.
\end{equation}
The sum runs over the first Brillouin zone and $a$ denotes the distance between adjacent $B$-sites. The resulting dispersion is given by
\begin{align}
\nonumber
\hbar\omega_{\pm}(k)&=
t \cos (ka)\pm\sqrt{t^2 \cos^2(ka)+2 t'^2 \cos (ka)+2t'^{2}}\\
&\stackrel{t'\rightarrow\sqrt{2}t}{=}
\begin{cases}
\quad-2t,\\
\quad\phantom{-}2t[1+\cos(ka)].
\end{cases}
\end{align}
We see that for the special ratio $\gamma=t'/t=\sqrt{2}$ we indeed obtain a flat lower band which is separated by an energy gap $\Delta=2t$ from the next band. The eigenstates take the form
\begin{align}
\label{eqn:stbloch}
\begin{pmatrix}
\beta_{-,k}\\
\beta_{+,k}
\end{pmatrix}
&=
\begin{pmatrix}
-\sin(\theta_{k}/2) & \cos(\theta_{k}/2)e^{-ika/2} \\
\cos(\theta_{k}/2) & \sin(\theta_{k}/2)e^{ika/2}
\end{pmatrix}
\vec b_{k},\\
\nonumber
\theta_{k} &=
\arctan
\left[
\frac{2\sqrt{1+\cos(ka)}}{\cos(ka)}
\right]+\pi\Theta[-\cos(ka)].
\end{align}
The flat dispersion allows to construct wave-packets of Bloch states (\ref{eqn:stbloch}) from the flat band that form localized  kinetic energy eigenstates. For example, we can obtain the states shown in Fig. \ref{fig:strictly} that are strictly localized to three sites, as
\begin{equation}
\ket{V_i}={\rm V}_{i}^{\dag}\ket{0} = \frac{1}{2}(\sqrt{2} b_{B,i}^{\dag}-b_{A,i-1}^{\dag}-b_{A,i}^{\dag})\ket{0}
\end{equation}

Because these states are strictly localized, one can use them to construct many-body ground states of the interacting Hamiltonian by occupying only nonoverlapping $\ket{V_i}$ states. In particular the CDW state on a lattice with $N$ sites is given by
\begin{equation}
|{\rm CDW}_{1/4}\rangle = \prod_{i=0}^{N/2} {\rm V}_{2i}^{\dag}|{\rm vac}\rangle.
\end{equation}
This state is clearly an eigenstate of the interaction energy with zero eigenvalue as there are no two sites with more than one particle. Note that $N/2$ particles correspond to a density $\nu_{c}=1/4$.

Although the states $\ket{V_i}$ are linearly independent and complete in the flat band subspace, they are not a convenient basis to work with because they are not orthogonal. This is evident from the fact that neighboring states $\ket{V_i}$ and $\ket{V_{i+1}}$ actually share a site.  We therefore define an alternative set of localized wave functions residing in the flat band, which do form an orthonormal basis. These are constructed as Wannier states.\cite{Kohn59,*Des-Cloizeaux63,*Des-Cloizeaux64,*Des-Cloizeaux64a}
\bea
W\yd_i\ket{0}&=&\sum_j \left(w_A^{*}(r_j-r_i)b\yd_{A,j}+w_B^{*}(r_j-r_i)b\yd_{B,j}\right)\ket{0},\nn\\
w_{B}(r_j) &=& -\int_{-\pi}^{\pi} \frac{dk}{2\pi} \sin(\theta_{k}/2) e^{ikr_j},\\
w_{A}(r_j) &=& \int_{-\pi}^{\pi} \frac{dk}{2\pi} \cos(\theta_{k}/2) e^{ikr_j-ik/2}.
\eea
By construction these states are orthogonal and exponentially localized around the unit cell $i$. Locally, they have the structure of the V-states but they fall off exponentially with a localization length of $\xi\approx \log(2.15)a$, cf. App.~\ref{app:projection}.

\subsection{Projection onto the flat band}
\label{sec:stproject}

The exact ground state $|{\rm CDW}_{1/4}\rangle$ at $\nu_{c}=1/4$ corresponds to a close-packing of V-states. If we dope the system with additional particles we loose this property and we expect a transition to a different phase. Let us assume small interactions $U<\Delta$. Any phase transition as a function of $\nu$ is then necessarily driven by the interactions as the kinetic energy is completely quenched for energies below $\Delta$. This provides strong motivation for projecting the interaction Hamiltonian onto the flat band, in analogy with projection to the lowest Landau-level for quantum Hall states.\cite{Girvin84}

To implement the projection we express the bosonic operators in terms of the Wannier basis derived above
\begin{equation}
b_{A(B),i}^{\dag}= \sum_{j} w_{A(B)}^{*}(r_{i}-r_{j}) W_{j}^{\dag} + \text{higher bands}.
\end{equation}
And project out the contributions from the higher band.  Due to the exponential tail of $w_{A(B)}(r)$ the projected Hamiltonian acquires arbitrarily long-ranged interaction terms. However, the rapid decay of the Wannier functions on the scale of a lattice constant allows us to truncate the sum to $|r_{i}-r_{j}|\leq a$.
\begin{figure}[b]
\begin{center}
\includegraphics{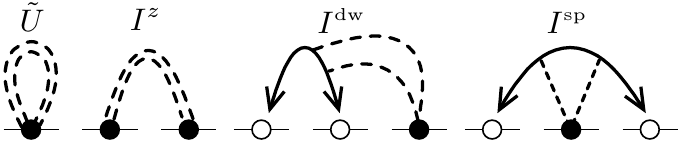}
\end{center}
\caption{
Processes in the spin-1/2 Hamiltonian (\ref{eqn:stspinham}). Arrows denote hopping (or spin-exchange) and dashed lines stand for density ($\sigma^{z}$) interactions.
}
\label{fig:stspinham}
\end{figure}

The resulting Hamiltonian contains a renormalized on-site interaction ($\tilde U$) and further range interactions as well as assisted hopping terms. Since all the coupling constants turn out to be smaller than the on site repulsion by a factor of $\sim\!4$ we replace $\tilde U$ with a hard core constraint. Now the effective Hamiltonian can be written as a spin-$1/2$ chain model:
\begin{align}
\nonumber
H_{\ssm eff} &= \sum_{i}\,
I^{z} \sigma_{i}^{z}\sigma_{i+1}^{z} +
\frac{I^{\ssm dw}}{2}
\left(  \sigma_{i}^{z}+1/2\right)\left(\sigma_{i\pm 1}^{+}\sigma_{i\pm 2}^{-} + {\rm H.c}\right)
\\
\label{eqn:stspinham}
&\qquad+ I^{\ssm sp}
\left[
\left( \sigma_{i}^{z}+1/2\right)\left(\sigma_{i-1}^{+}\sigma_{i+1}^{-} + {\rm H.c}\right)
\right].
\end{align}
The interaction constants derive from the overlap of Wannier functions on different sites and are given by (see App.~\ref{app:projection} for more details)
\begin{align}
\tilde U & \approx \phantom{-}0.40\, U \;(\rightarrow \infty), &I^{z}& \approx \phantom{-}0.112 \,U, \\ I^{\ssm dw}&\approx  -0.025\,U,
&I^{\ssm sp}& \approx  -0.011\, U.
\label{eqn:stparam}
\end{align}

Note that all terms are proportional to the interaction $U$. This is because the effective Hamiltonian (\ref{eqn:stspinham}) is just the projection of the interaction onto the flat band. For the same reason all terms in the Hamiltonian are four--$W$ operator terms. Even off-diagonal terms exclusively include {\em assisted} hopping $\propto I^{{\ssm dw}({\ssm sp})}$. The processes in Hamiltonian (\ref{eqn:stspinham}) are illustrated in Fig.~\ref{fig:stspinham}.

Note that Hamiltonian (\ref{eqn:stspinham}) describes a spin system on a linear chain. The sawtooth geometry is absorbed by the projection onto the flat band. The density $\nu_{c}$ of the original model now translates to half-filling, or zero magnetization of the effective spin-1/2 model. Furthermore, the exact CDW solution of the underlying model corresponds to a perfect N\'eel order of the spin chain. Due to the truncation of the Wannier wave functions, this state is not an exact eigenstate anymore. However, the dominating Ising interaction $I^{z}\gg I^{{\ssm dw}({\ssm sp})}$ still leads to a large staggered magnetization in the ground state at zero magnetization.

\subsection{Analysis of the effective Hamiltonian}
\label{sec:stheff}

We are now in the position to investigate the fate of the CDW at lattice filling above $\nu_{c}$, which corresponds to $\langle \sigma^z \rangle>0$ in the effective model.  The Hamiltonian (\ref{eqn:stspinham}) is similar to a $X\!X\!Z$-model with a dominating Ising interaction, but with hopping terms replaced by assisted hopping. In the latter model the charge density is destroyed upon doping away from the magic filling by the mechanism known as the commensurate--incommensurate transition. \cite{Pokrovsky79,Schulz80,Chitra97} We will show that the system described by Hamiltonian (\ref{eqn:stspinham}) undergoes the same transition.

As usual in one dimension single particles doped into the commensurate CDW can either go into an interstitial site of the CDW or break up into a pair of domain--walls in that order, each carrying half of the added charge.\cite{Sachdev99} We will show below, that  energetically, condensation of domain walls is favorable to that of the interstitial sites. With this input we shall derive the appropriate long wavelength description of (\ref{eqn:stspinham}) via bosonization, which yields specific predictions for the emergent critical state at $\nu > \nu_{c}$.

\subsubsection{Domain walls versus interstitial condensate}

Particles added to the CDW state correspond to adding up-spins to the effective spin model on top of the $z$-Ne\'el state. The flipped spins can behave in two ways: (i) they can exist as well defined excitations and condense as interstitials without destroying the CDW order; or (ii) break up into two mobile domain walls in the Ising spin configuration, each carrying a spin-$1/2$, thus leading to destruction of the CDW order. If one thinks of domain walls as the elementary excitations, then the option (i) above can be viewed as having a bound state of a kink and an anti-kink. Which of the two options is realized in a specific model is ultimately a matter of energetics.

Let us then estimate and compare the energies of an interstitial to that of a pair of free domain walls. The Ising interaction energy associated with an interstitial (flipped spin) is $2I^{z}$ because this state violates two Ising bonds. In addition, the interstitial can hop to the next empty site (down-spin) via the term $I^{\ssm sp}$. Therefore, the kinetic energy gain is $-2I^{\ssm sp}$.

The pair of domain walls still violate two bonds and therefore incur the same interaction energy as the interstitial, i.e., $2I^z$. However, each domain wall can move due to the action of the assisted hopping term $I^{\ssm dw}$, which leads to the kinetic energy gain  $-4I^{\ssm dw}$  per added particle (two domain walls).  With the coupling constants given by (\ref{eqn:stparam}), the pair of domain walls are seen to be clearly favorable to the interstitial. We therefore conclude that the long-range order of the CDW state at $1/4$ filling will be immediately destroyed by proliferation of free domain walls upon increasing the density.

\subsubsection{Long wavelength description}

The energetic considerations detailed above imply destruction of the density wave state immediately upon increasing the density above $\nu_c=1/4$. The emergent critical state is naturally described within the continuum limit by bosonizing the effective spin-chain model (\ref{eqn:stspinham}). However, the usual scheme,\cite{Giamarchi04} of first mapping to a fermion system using a Jordan-Wigner transformation is not straightforward. Because there are no quadratic terms in the spin-1/2 Hamiltonian, there results no obvious Fermi surface to support an expansion in slow modes. Under such conditions it is more convenient to employ the so called phenomenological approach to bosonization of bosons.\cite{Haldane82} For this, we reintroduce the on-site interaction $\tilde U$ and write the density and field operators as
\begin{align*}
\rho_{i}&=
\left[
\frac{1}{2a}+\frac{\partial_{x}\theta(x_{i})}{\pi}
\right]
\sum_{m\in \mathbb{Z}}e^{i[\pi m-2m\theta(x_{i})]},
\\
W_{i}&=e^{i\phi(x_{i})}
\left[
\frac{1}{2a}+\frac{\partial_{x}\theta(x_{i})}{\pi}
\right]^{1/2}
\sum_{m\in \mathbb{Z}}e^{i[\pi m-2m\theta(x_{i})]}.
\end{align*}
Note, that we expand around $\nu=1/2a$. The most relevant density wave captured with this approach is of wavelength $\lambda=2a$ and is described by the order parameter $\langle \cos(4\theta(x))\rangle$. We take a naive continuum limit by rewriting (\ref{eqn:stspinham}) in terms of the new fields $\phi(x)$ and $\theta(x)$, keeping only the most relevant terms we obtain
\begin{equation}
\label{eqn:stharmonic}
H_{\ssm c}=\frac{u}{2\pi}\int\! dx
\left[
\frac{1}{K}(\partial_{x}\theta)^{2}+K(\partial_{x}\phi)^{2}
\right]
+
\frac{\Omega}{2\pi}\int\! dx  \cos(4\theta),
\end{equation}
with
\begin{align*}
u&=\frac{2a}{\pi }\sqrt{[\tilde U-3(I^{\ssm dw}+I^{\ssm sp})](I^{\ssm dw}/4+I^{\ssm sp})}\approx \frac{Ua}{\pi}\,0.08,\\
K&=\sqrt{\frac{I^{\ssm dw}/4+I^{\ssm sp}}{\tilde U-3(I^{\ssm dw}+I^{\ssm sp})}}
\approx 0.42,\\
\Omega &=
\frac{\pi}{a}(\tilde U-I^{z}-I^{\ssm dw}-I^{\ssm sp})\approx \frac{U\pi}{a}\, 0.05.
\end{align*}
We find the expected CDW reproduced also within this long-wavelength theory, as the term $\propto \cos(4\theta)$ is relevant for $K<1/2$. Note, that instead of taking the naive continuum limit one can regard (\ref{eqn:stharmonic}) as a phenomenological description of the CDW, as long as $K<1/2$. The results we draw from (\ref{eqn:stharmonic}) will not depend on these details but rely solely on the form of the cosine term.

\begin{figure}[tb]
\begin{center}
\includegraphics{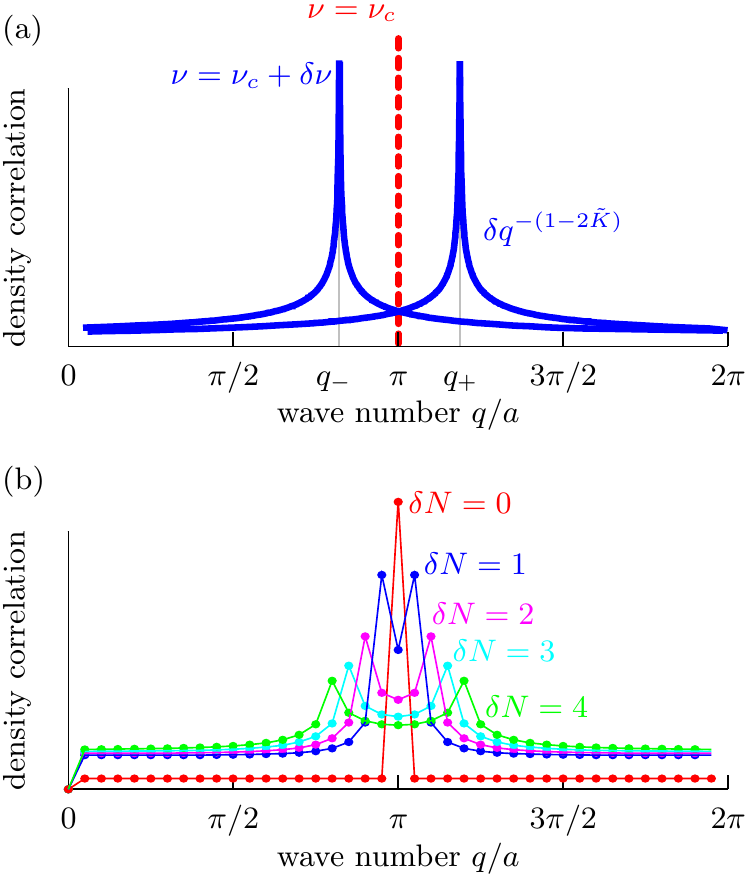}
\end{center}
\caption[Analytic density correlations]{
(color online). Fourier transform of the density correlations in the \st chain. (a) Prediction of the long wavelength theory. The delta function peak at  $q=\pi/a$ (red dashed line), which signals the CDW at $\nu=\nu_{c}$, splits with increased density $\nu=\nu_c+\d\nu$ into two power-law peaks (blue full line)at $q_{\pm}=\pi(1\pm 2\delta\nu)$. The effective Luttinger parameter, which controls the power-law is given in Eq. (\ref{eqn:stK}). (b) Density correlations computed using DMRG. The splitting of the peak by one quanta $\d k={2\pi\over L}$ per added particle confirms the predicted $q_{\pm}$ of the long-wavelength theory.
}
\label{fig:correlations}
\end{figure}

Let us now proceed with the discussion of an additional density $\delta\nu\!=\!\nu\!-\!\nu_{c}$ above the CDW. Forcing $\delta\nu>0$ is equivalent to a chemical potential larger then the CDW gap $\Delta_{\ssm CDW}$. To characterize the resulting gapless state above $\Delta_{\ssm CDW}$,  we can use results from the commensurate--incommensurate transition introduced by Pokrovsky and Talapov,\cite{Pokrovsky79} in particular the critical behavior discussed by Schulz.\cite{Schulz80} The long-wavelength theory is now given by a Luttinger liquid with \cite{Chitra97}
\begin{equation}
\tilde K=\frac{1}{4}(1+\mathcal{D} \delta\nu),\label{eqn:stK}
\end{equation}
where $\mathcal{D}$ is a positive number independent of $\delta\nu$. Furthermore, we can calculate correlation functions by using the new Luttinger liquid with $\tilde K$ and by introducing a shift $\theta(x)\rightarrow \theta(x)+\pi x \delta\nu $. One readily obtains  \cite{Chitra97,Giamarchi04}
\begin{align}
\langle \rho(x)\rho(0)\rangle_{q\approx \pi/a}&\approx
\cos(\pi(1+2\delta\nu)x/a)\left(\frac{a}{x}\right)^{2\tilde K}, \\
\langle W^{\dag}(x)W(0)\rangle_{q\approx 0} &\approx
\left(
\frac{a}{x}
\right)^{1/2\tilde K}.
\end{align}
We show a sketch of the density correlations in Fig.~\ref{fig:correlations}(a), together with the $\nu=\nu_{c}$ result.

From the above considerations we conclude that under particle doping the exact ground state $|{\rm CDW}_{1/4}\rangle$ melts due to delocalization of domain walls. The harmonic fluid approach allows for a prediction of both the functional form of the density--density correlations and the effective Luttinger liquid parameter $\tilde K$. In the next section we test these predictions by numerical analysis of the \st chain.

\subsection{Numerical analysis}
\label{sec:stDMRG}

To complement the analytical study and confirm its predictions we analyze the original \st lattice model numerically using the DMRG method. Calculations are performed on an 80 site lattice ($L=40$ unit cells) with the parameters $t=U=1$ and $t'=\sqrt{2}$ and we retain up to 200 states. In all calculations we make use of the ALPS DMRG code.\cite{alps-url,*Albuquerque06}  We consider particle numbers $N=20+\d N$ ($\d N =0,1,2,3,4$) particles. Note, that $\d N=0$ corresponds to $\nu=\nu_{c}$, and we indeed find the exact ground-state $|{\rm CDW}_{1/4}\rangle$ at this point. This is evident from the perfect staggered density correlations seen in Fig.~\ref{fig:correlations}(b) as a sharp peak in the static structure factor at wave-vector $\pi/a$. As we dope the system above $\nu_c$ the peak splits in two, with a shift that grows by $2\pi/L$ per added particle. Exactly as expected from the bosonization study above.

\begin{figure}[tb]
\begin{center}
\includegraphics{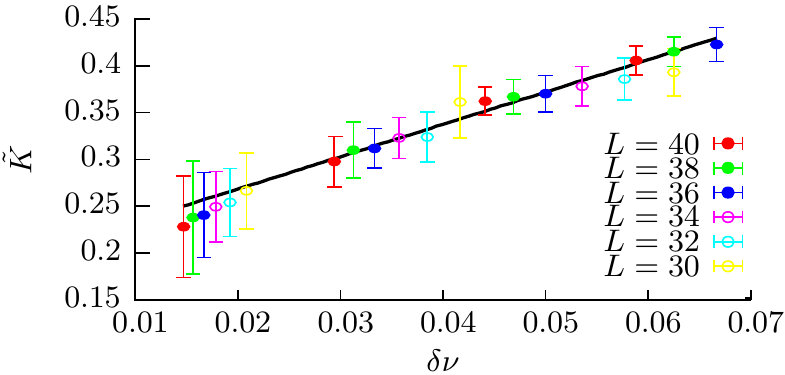}
\end{center}
\caption{
(color online) The effective Luttinger parameter $\tilde K$ as a function of density $\delta\nu=\nu-\nu_{c}$ extracted from the off-diagonal correlation function calculated using DMRG. The black line is a guide to the eye.
}
\label{fig:keff}
\end{figure}

To capture the critical behavior of the system it is better to extract the off-diagonal correlations, because they have a faster decay. At the special point $\nu=\nu_{c}$ these are found to be strictly local, as expected. For $\d N>0$ on the other hand the decay of the off-diagonal correlations is consistent with a power-law. As the density is increased the critical exponent is reduced continuously from $\a\approx 2$ at $\d N=0^+$. This is consistent with a Luttinger parameter $\tilde K$ which grows continuously from $\tilde K \approx 1/4$, as predicted by the long-wavelength theory above, cf. Fig.~\ref{fig:keff}.

The numerical study thus confirms our main predictions. The density wave state is immediately destroyed upon increase of the density. The exact density wave state is replaced by power-law density correlations at the incommensurate wave-vector consistent with the doping away from $\nu_c$. The emergent Luttinger liquid phase that forms is characterized by a Luttinger parameter $\tilde K\gtrsim 1/4$.

\section{\kg lattice}
\label{sec:kagome}

We now turn to the discussion of the \kg lattice, focusing on the question, which ground state forms when the density is increased beyond $\nu_c=1/9$.

\subsection{Flat band} \label{sec:kgflat}

Let us review the essential features of the tight-binding Hamiltonian on this lattice.  It is useful to introduce the vector notation
\begin{equation}
\nonumber
\vec b_{i}=
\begin{pmatrix}
b_{i,A}\\
b_{i,B}\\
b_{i,C}
\end{pmatrix}
\quad
\mbox{and}
\quad
\vec b_{\bf k} =
\begin{pmatrix}
b_{{\bf k},A}\\
b_{{\bf k},B}\\
b_{{\bf k},C}
\end{pmatrix} =
\frac{1}{\sqrt{M}}\sum_{i} e^{i{\bf k}\cdot {\bf R}_{i}}
\vec b_{i}.
\end{equation}
The unit cells $i$ are at positions ${\bf R}_{i}=m_{i}{\bf a}_{1}+n_{i}{\bf a}_{2}$, $m_{i},n_{i}\in \mathbb{Z}$ with ${\bf a}_{1}=a(1,0)$, ${\bf a}_{2}=a(1/2,\sqrt{3}/2)$, and $a$ the lattice constant. Furthermore, the subscripts $A$, $B$, and $C$ label the lattice sites in a unit cell at ${\bf R}_{i}$, ${\bf R}_{i}+{\bf a}_{1}/2$, and ${\bf R}_{i}+{\bf a}_{2}/2$, respectively, cf. Fig.~\ref{fig:models}(b); $M$ denotes the total number of unit cells. The wave vectors are confined to the first Brillouin zone as defined by the Wigner--Seitz cell with respect to the reciprocal lattice vectors ${\bf Q}_{1}=\frac{2\pi}{a}(1,-1/\sqrt{3})$ and ${\bf Q}_{2}=\frac{2\pi}{a}(0,2/\sqrt{3})$, see insets of  Fig.~\ref{fig:models}(a). The hopping Hamiltonian now reads
\begin{equation}
H=|t|\sum_{\bf k}\vec b_{\bf k}^{\dag}
\begin{pmatrix}
0 & 1+e^{ik_{1}} & 1+e^{ik_{2}} \\
1+e^{-ik_{1}} & 0 &  1+e^{ik_{3}} \\
1+e^{-ik_{2}} & 1+e^{-ik_{3}} & 0
\end{pmatrix}
\vec b_{\bf k},
\end{equation}
where $k_{\nu}={\bf k}\cdot {\bf a}_{\nu}$, $\nu=1,2$ and $k_{3}=k_{1}-k_{2}$. Diagonalizing this matrix gives the three bands:
\begin{align}
\nonumber
\hbar\omega_{0}({\bf k})&=-2t,\;\;
\hbar\omega_{\pm}({\bf k})=t\bigl[1\pm\sqrt{3+2\Lambda({\bf k})}\bigr],\,\,
\mbox{with} \\ \Lambda({\bf k})&=\cos(k_{1})+\cos(k_{3})+\cos(k_{2}),
\end{align}
shown along the high symmetry lines in Fig.~\ref{fig:models}(a). The eigenstates corresponding to the flat band are given by
\begin{equation}
\langle {\bf k}|\beta_{0,{\bf k}}\rangle=
\sqrt{\frac{2}{-3+\Lambda({\bf k})}}
\begin{pmatrix}
\sin(k_{3}) \\ \phantom{-}e^{-ik_{1}/2}\sin(k_{2}) \\ -e^{-ik_{2}/2} \sin(k_{1})
\end{pmatrix}.
\end{equation}

The existence of a flat band allows to construct localized eigenstates of the same energy. The most local wave-function of this type is confined within a single hexagon\cite{Bergman08}
\begin{equation}
\label{eqn:hexagon}
A_{i}^{\dag} = \frac{1}{\sqrt{6}} \sum_{\alpha_{i}=1}^{6} (-1)^{\alpha_{i}} b_{\alpha_{i}}^{\dag}.
\end{equation}
Here $\alpha_{i}$ labels the sites around the hexagon, cf. Fig.~\ref{fig:strictly}(a) and $i$ labels the hexagon.  It is easy to see that destructive interference on the sites adjacent to the hexagon sites prevent the spread of the wave-function. Delocalization on the hexagon gives rise to the kinetic energy $-2t$.  Therefore $A_{i}^{\dag}$ indeed creates an eigenstate with the energy of the flat band. Moreover, by acting on the vacuum with the operators $A_{i}^{\dag}$ to create any number of spatially {\em nonoverlapping} states, it is possible to the construct exact many-body ground-states of the total interacting Hamiltonian (\ref{eqn:bosonic-hamiltonian}). For sufficiently low lattice filling there is a huge number of degenerate arrangements of the localized states. However, these localized orbitals become close packed at the critical filling $\nu_{c}=1/9$, at which point there are only three ground states corresponding to the three degenerate density wave states:
\begin{equation}
\label{eqn:kgexact}
|{\rm CDW}_{1/9}\rangle = \prod_{i \in \mathbb{T}} A_{i}^{\dag} | {\rm vac}\rangle.
\end{equation}
The set $\mathbb{T}$ contains close packed nonoverlapping hexagons as depicted in Fig.~\ref{fig:strictly}(c).

At higher lattice filling $\nu>1/9$ such a construction would fail to create eigenstates of the interaction term because it is no longer possible to construct a spatially nonoverlapping set of states.  Our strategy to address this regime is, the same as in the \st model, to derive an effective Hamiltonian using a projection on the flat band.  For this purpose we shall introduce an orthonormal set of localized states, which belong exclusively to the flat band, and are defined on every unit cell. These are the Wannier states
\begin{equation}
\label{eqn:kgwannier}
\vec W_{i}({\bf r})= \int\frac{d{\bf k}}{v_{0}}\,\langle {\bf k}|\beta_{0,{\bf k}}\rangle e^{i{\bf k}\cdot ({\bf r}-{\bf r}_{i})},
\end{equation}
where the integral runs over the first Brillouin zone and $v_{0}$ denotes its volume. Note, that due to the band-touching at the $\Gamma$-point, these Wannier states are not exponentially localized but have power-law tails.\cite{Des-Cloizeaux63} Their slowest decay is along high-symmetry directions of the lattice where they fall off $\vec W({\bf r})\sim 1/|{\bf r}|$; see App.~\ref{app:projection} for more details.

\subsection{Projection into the flat band}
\label{sec:kgproject}

We are now in the position to derive an effective Hamiltonian for the low-energy dynamics by projecting the interactions onto the flat band. In this case the projection is fraught with further complication because of the vanishing gap  in the single particle spectrum between the flat band and the second Bloch band.  Recall that in the case of the \st model the projection was controlled by the band gap $\Delta=2t$, which gave rise to the small parameter  $U/\Delta$. The situation is different on the \kg lattice because the non-interacting model has a topologically protected\cite{Bergman08,Green10} band touching ($\Delta\!=\!0$) at the $\Gamma$ point.  Below we argue that for filling factors $\nu\gtrsim 1/9$ the interactions open an effective band gap which ultimately controls the projection.\cite{Zhitomirsky05}

The key point is that a particle in the second Bloch band, with small quasi-momentum $k\approx 0$, suffers an energy shift $\D \approx U/9$ due to interaction with the mean density $\av{\nu}_i$. A particle on the flat band on the other hand, does not suffer the same mean-field energy shift, because of the extra freedom it has to be in superpositions of Bloch states. This is explicitly evident in the energy scales of the effective spin Hamiltonian generated by projection to the flat band, which turn out to be $\sim 0.02U$, much less than the energy shift $\D\approx U/9$ of the second band. Thus the projection is justified a posteriori.

Technically we carry out the projection by expressing the boson operators appearing in the original Hamiltonian (\ref{eqn:bosonic-hamiltonian}) using the Wannier operators $W_{i}$ while truncating the contribution from Wannier states of the higher bands. The resulting Hamiltonian then contains terms of the form $I_{ijrs}W_{i}^{\dag}W_{j}^{\dag}W_{r}W_{s}$, where the coefficients $I_{ijrs}$ consist of wave-function overlaps of the Wannier functions located at $i,j,r,$ and $s$. The largest term is an effective on-site interaction $\tilde U \approx 0.14 \, U$, which comes from $i=j=r=s$. The next coefficient stems from $i=j$ and $r=s$ with $i$ and $r$ nearest neighbors. The resulting nearest-neighbor interaction has magnitude $I_{z}\approx 0.02\,U\ll \tilde U$. Other terms, as we shall see below are even smaller. We can therefore safely replace $\tilde U$ with a hard core constraint. This restriction is also required in order to be consistent with neglecting occupation of particles in the second band, which is at an energy $\D\approx U/9$ above the low-energy manifold, but comparable to the on-site interaction. The projected Hamiltonian is therefore given by a hard-core boson (or spin-1/2) model on a triangular lattice.

The effective Hamiltonian contains hopping and interaction terms which fall off with range as $1/|{\bf r}|^{2}$, or faster, due to the algebraic decay of the Wannier functions. To keep the model simple, we truncate all interaction terms smaller than $I_{\ssm cut}=I^{z}/20$. When necessary, we check against convergence with respect to this cut-off.

\begin{figure}[b]
\includegraphics{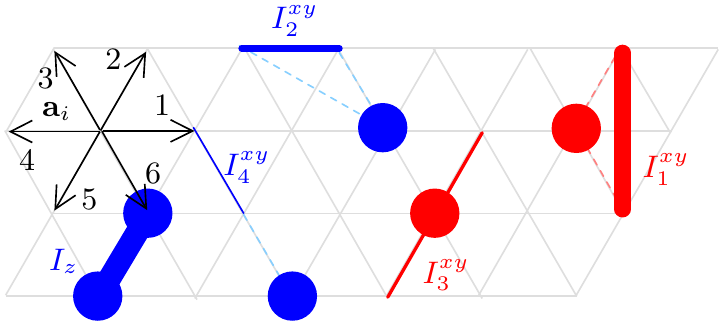}
\caption{
(color online) Illustration of the processes appearing in the effective Hamiltonian (\ref{eqn:kgeff}) for the \kg lattice. A dot represents a density operator, lines denote either a nearest neighbor interaction ($I^{z}$) or a hopping ($I^{xy}$) assisted by the nearby density operator. Note that $I^{xy}_{1/3}$ have negative effective hopping amplitudes (red) while $I^{xy}_{2/4}$ have positive hopping (blue).
}
\label{fig:assisted}
\end{figure}

To explicitly write the effective Hamiltonian we define the vectors ${\bf a}_{\alpha}$, $\alpha=1,..,6$, connecting the nearest neighbors on the triangular lattice as shown in Fig.~\ref{fig:assisted}. The Hamiltonian in terms of spin-1/2 operators $\sigma^{\pm}_{i}$, $\sigma^{z}_{i}$ is given by
\begin{widetext}
\begin{align}
\nonumber
H_{\ssm eff}&\approx\sum_{i}\sum_{\alpha=1}^{6}\biggl[ \frac{I^{z}}{2}
 \sigma_{{\bf r}_{i}}^{z}\sigma_{{\bf r}_{i}+{\bf a}_{\alpha}}^{z}
+(\sigma_{{\bf r}_{i}}^{z}+1/2)
\biggl\{
I_{1}^{xy}
\sigma_{{\bf r}_{i}+{\bf a}_{\alpha}}^{+}\sigma_{{\bf r}_{i}+{\bf a}_{\alpha+2}}^{-}
+
I_{2}^{xy}
\sigma_{{\bf r}_{i}+{\bf a}_{\alpha}}^{+}
(
\sigma_{{\bf r}_{i}+{\bf a}_{\alpha}+{\bf a}_{\alpha+1}}^{-}+
\sigma_{{\bf r}_{i}+{\bf a}_{\alpha}+{\bf a}_{\alpha-1}}^{-}
)
\\
\label{eqn:kgeff}
&\qquad\qquad\qquad\qquad\qquad\qquad\qquad\qquad
+
I_{3}^{xy}
\sigma_{{\bf r}_{i}+{\bf a}_{\alpha}}^{+}\sigma_{{\bf r}_{i}+{\bf a}_{\alpha+3}}^{-}
+
I_{4}^{xy}
\sigma_{{\bf r}_{i}+{\bf a}_{\alpha}}^{+}\sigma_{{\bf r}_{i}+2{\bf a}_{\alpha}}^{-}
+{\rm H.c.}
\biggr\}+\dots
\end{align}
\end{widetext}
The coefficients read
\begin{align}
\nonumber
I^{z} & \approx \phantom{-}0.028\,U &I^{xy}_{1}/I^{z}& \approx -1/3,
&I^{xy}_{2}/I^{z}&\approx  1/6, \\
I^{xy}_{3}/I^{z}& \approx  -1/8,
&I^{xy}_{4}/I^{z}& \approx 1/16.
\label{eqn:kgparam}
\end{align}
In Fig.~\ref{fig:assisted} we illustrate the processes in $H_{\ssm eff}$.

Note that by projecting to a single band we removed $2/3$ of the degrees of freedom and thus changed the lattice structure to a triangular lattice with one instead of three sites per unit cell. The critical density at which the density wave states form in the \kg lattice is translated accordingly to $1/3$ filling of the effective model on the triangular lattice. Because the repulsive (or Ising antiferromagnetic) interaction $I_z$ is the largest energy scale in (\ref{eqn:kgeff}) we do indeed expect the density-wave state to form at this filling. Note, however that the state with {\em exactly} one boson on every three sites is not an exact eigenstate of the effective model (\ref{eqn:kgeff}). This is of course due to the truncation of longer range interactions in (\ref{eqn:kgeff}).  In what follows we will study the properties of the effective model and the fate of the CDW order for filling $\nu\gtrsim 1/3$ of the triangular lattice.

\subsection{Domain walls versus interstitial condensate}

When we increase the lattice filling beyond $\nu_c=1/9$ of the \kg lattice or $1/3$ of the effective triangular lattice we necessarily introduce defects to the density wave structure with one particle in every three unit cells.  There are different types of defects that can accommodate the extra particles. On the one hand, the extra particles can hop around interstitial sites of the lattice, which is innocuous to the underlying density wave structure. On the other hand, they can nucleate mobile domain walls in the CDW order, that ultimately destroy it.

An extra particle in an interstitial site has three nearest neighbors in the CDW background. The resulting interaction energy per particle is $3I_{z}$. The particle can also hop between interstitial sites and thereby gain kinetic energy from the assisted hopping matrix elements $I^{xy}_{1}$ and $I_{3}^{xy}$. The delocalization energy of the added particle is $-6|I_{1}^{xy}|-3 |I_{2}^{xy}|\approx -2.5 I_z$.

\begin{figure}[t]
\includegraphics{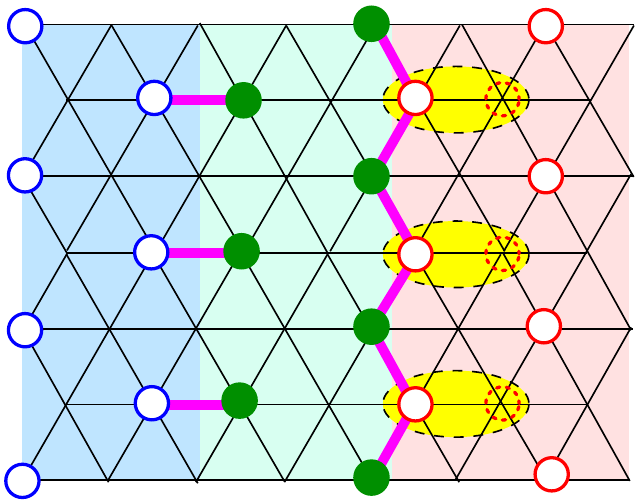}
\caption{
(color online) Three patches of the CDW (different colors/gray scale) with two types of domain walls: "straight" (left) and "zigzag" (right). The thick lines denote the violated bonds which lead to the interaction energy cost. The yellow (shaded) ellipses around half of the particles in the ``zigzag" domain wall indicate resonating particles which gain the delocalization $I^{xy}_2$.
}
\label{fig:strip}
\end{figure}

Let us compare to the case when particles are added as domain walls. Domain walls between the three degenerate CDW states come in two flavors, the ``straight'' and ``zigzag'' domain walls, which are illustrated in Fig.~\ref{fig:strip}. The line density, or number of added particles, associated with the ``zigzag'' domain wall is twice that of the ``straight" domain wall. However, they both incur the interaction energy cost of $3I_z$ per added particle, which is the same as the interaction energy of a particle in an interstitial site.

The differences in energy between the defects stems from the delocalization energy of particles within them. Consider first the ``zigzag'' domain wall. Half of the particles in it can resonate independently on the horizontal bonds, as shown by the ellipses in Fig. \ref{fig:strip}, facilitated by the assisted hopping process $I_2^{xy}$. The energy gain per added particle is then seen to be $-6|I_2^{xy}|\approx I_z$.  Particles in the "straight" domain wall, on the other hand, can hop with the matrix-element $I_2^{xy}$ only to reach a point where the interaction energy is increased by one bond $I_z$. Therefore, they can only move coherently by a second order process, gaining them delocalization energy of order $(I_2^{xy})^2/I_z$. In both cases, of ``zigzag" and ``straight" domain walls, the energy gain from delocalization is less than that of the interstitial defect.

We conclude that doping of particles into interstitial sites of the CDW, due to the high mobility they acquire, is energetically favorable to nucleating domain walls. Since the interstitial particles do not destroy the CDW, we expect that order to survive and co-exist with a condensate formed by the mobile interstitial particles. In the next section we use the effective spin model on the triangular lattice to formulate a mean-field description of the resulting supersolid phase.

\label{sec:kgdomain}
\subsection{Mean-field treatment}
\label{sec:kgmean}

In the last section we established that the added particles preferably hop on interstitial sites of the CDW and do not lead to proliferation of domain walls. Consequently the CDW order is preserved and a mean-field treatment of the coexisting CDW and interstitial condensates is justified. The mean-field approximation consists of making a site factorizable ansatz for the wave-function:
\begin{equation}
\label{eqn:kgmeanfield}
|\Psi\rangle =\prod_{i}
[\cos(\theta_{i}/2)|\downarrow\rangle_{i}+e^{i\phi_{i}}\sin(\theta_{i}/2)|\uparrow\rangle_{i}],
\end{equation}
In order to account for the broken translational symmetry, we assume space dependent coefficients $\theta_{i}$ and $\phi_{i}$ with the unit cell of the CDW at filling $\nu_c=1/9$. The variational energy $E_{\ssm var}=\langle \Psi |H_{\ssm eff} |\Psi\rangle$, then depends on the three angles $\theta_{\alpha}$ and phases $\phi_{\alpha}$ as
\begin{figure}[t]
\includegraphics{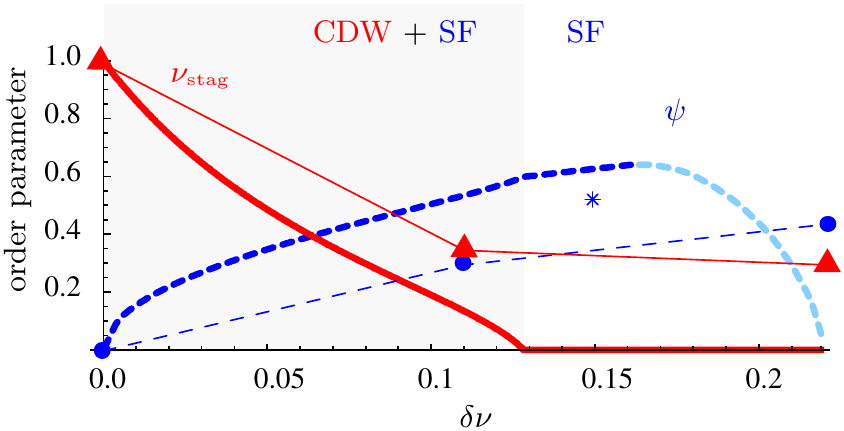}
\caption{
(color online) Order parameters in the \kg system as a function of the density $\nu=\nu_c+\d\nu$. The lines are the result of the mean-field analysis of the effective Hamiltonian on the flat band. The CDW order parameter is shown as a full (red) line while the superfluid order parameter is the dashed (blue) line. Circles (CDW) and triangles (superfluid) mark the values obtained from exact diagonalization of the original \kg system on a small cluster. The (blue) star at $\delta\nu=0.15$ is the fluctuation reduced superfluid order parameter. The mean-field result shows a critical density where the CDW order vanishes.
}
\label{fig:mf}
\end{figure}
\begin{align}
\nn
&\;E_{\ssm var}[{\t_\a,\f_\a}]=\sum_{\rm c.r.}\,\biggl\{
-|I_{1}^{xy}|
[\sin(\theta_{A})+ \sin(\theta_{B})]\sin^{2}\frac{\theta_{C}}{2}
\\
\nn
&\;
+|I_{2}^{xy}|\cos(\phi_{AB})
\left[\sin^{2}\!\frac{\theta_{A}}{2}+ \sin^{2}\!\frac{\theta_{B}}{2}\right]  \sin\frac{\theta_{A}}{2}\sin\frac{\theta_{B}}{2}
\\
&\;
+4I_{z} \sin^{2}\!\frac{\theta_{A}}{2}\sin^{2}\!\!\frac{\theta_{B}}{2}\biggr\}.
\label{eqn:mfenergy}
\end{align}
The sum runs over cyclic permutations of $A$, $B$, and $C$ and $\phi_{AB}=\phi_{A}-\phi_{B}$, etc. Here we only gave the explicit expression for the expectation values of the leading terms in (\ref{eqn:kgeff}). When solving the variational problem numerically, we also included, and found no significant effect, of the other terms.

Let us first consider two simple limits of the variational problem. The pure CDW state corresponds to the configuration $\theta_{A}=\pi$ and $\theta_{B}=\theta_{C}=0$.  Plugging this into (\ref{eqn:mfenergy}) we see that all terms,``interaction" and ``kinetic", vanish and the phases $\f_\a$ all drop out of the problem. The other simple limit is a pure condensate with no density modulation, so that $\theta_{A}=\theta_{B}=\theta_{C}\equiv\theta$. In this case, the problem reduces to
\begin{multline*}
E[{\theta,\f_\a}]=
2\sin^2\frac{\theta}{2} \biggl\{(-3|I_{1}^{xy}| \sin \theta+
3 I_{z}(1-\cos (\theta))
\\
+|I_{2}^{xy}| \sin\theta
\bigl[
\cos (\phi_{AB})+\cos (\phi_{BC})+\cos (\phi_{CA})
\bigr]
\biggr\}.
\end{multline*}
The value of $\theta$ is set by the density $\delta\nu$ and we are left with an optimization with respect to the phases $\phi_{\alpha}$. The energy is minimal for
\begin{equation}
\phi_{A} = 0, \quad \phi_{B} = \pm 2\pi/3, \quad \phi_{C} = \pm 4\pi/3.
\end{equation}
For the experimental realizations discussed below, it is interesting to note that when Fourier transformed, this phase pattern corresponds to the $K$ or $K'$ points of the Brillouin zone of the triangular lattice. However, in the reduced CDW Brillouin zone these points correspond to the $\Gamma$ point.

In the general case we optimize for the angles $\theta_{\alpha}$ and phases $\phi_{\alpha}$ numerically to find the mean-field density pattern. Starting from an initial guess of the form $\theta_{A}=\pi$, $\theta_{B/C}=0$, the minimal energy is always obtained for a solution of the form $\theta_{A}\leq \pi$, $\theta_{B}=\theta_{C}\geq 0$. Furthermore, above the critical filling (i.e.  $\delta\nu>0$) the phases quickly approach the $120^{\circ}$ pattern found above for the case of a pure condensate.

Using the optimized variational wave-function we compute the order parameters of the CDW
\be
\nu_{\ssm stag}=
\frac{\langle \Psi|
\sigma_{A}^{z}-\sigma_{B}^{z}|\Psi\rangle
}{\sum_{\alpha}\langle \Psi|
\sigma_{\alpha}^{z}| \Psi\rangle}.
\ee
and of the superfluid
\be
\qquad
\psi=|\langle \Psi | \sigma^{+}_{B} | \Psi \rangle|.
\end{equation}
Both are shown in Fig.~\ref{fig:mf}. The CDW order parameter $\nu_{\ssm stag}$ is seen to decrease continuously from the perfect CDW at the magic filling $\nu_c=1/9$ ($\delta\nu=0$) until it vanishes at an upper critical density $\delta\nu_{\ssm MF}\approx 0.13$. At the same time there is a build up of the superfluid order parameter $\psi$.  The change from an increase to a reduction and eventual vanishing of $\psi$ with density at $\nu=1/3$ is an artifact of the mapping to spin-$1/2$ degrees of freedom (hard core bosons). However at this stage our lowest band approximation breaks down. The energy cost of interaction with neighboring particles grows to a point that it becomes energetically favorable to occupy the second band (and consequently also induce double occupancy).

The above analysis of the effective model should give qualitatively correct results at moderate densities, but should not be relied upon for high quantitative precision. One cause for quantitative error is the truncation of further than nearest neighbor interactions that were generated by the projection to the lowest Bloch band. Second, the mean field approximation neglects quantum fluctuations of the effective spin-$\half$ degrees of freedom. Inclusion of these effects certainly lead to quantitative changes, such as a shift of the upper critical filling $\delta\nu_{\ssm MF}$ at which the CDW order vanishes. Nevertheless the main features, including the existence of a finite window of coexisting CDW and superfluid orders, as well as the ordering pattern of the superfluid, are expected to be robust.

We have quantified the effect of quantum fluctuations, which were neglected in the mean-field approximation by including spin-wave corrections. This is done using the Holstein-Primakoff approach (see for example Ref.~\onlinecite{Auerbach94}), where we expand the Hamiltonian to quadratic order in the Holstein-Primakoff bosons. The reduced value of the superfluid order parameter due to this effect is shown as a star in Fig. \ref{fig:mf} for one point in the phase diagram (at $\delta\nu=0.15$).

\subsection{Exact diagonalization}
\label{sec:kged}

To support the above predictions, in particular the existence of a supersolid phase at densities $\nu>1/9$, we carry out an exact diagonalization (ED) study of the original \kg lattice model (\ref{eqn:bosonic-hamiltonian}) on finite clusters.

What are we looking for?  We know from exact considerations, that at very low density $\nu<1/9$ the system is infinitely compressible since filling in nonoverlapping localized flat band states does not incur an interaction energy cost. The effective theory and in particular the mean-field results imply that above the magic filling $\nu>1/9$ one obtains a compressible state. Doping extra particles into interstitial sites of the CDW state incurs a non-vanishing interaction energy cost. This prediction is not obvious and is worth verifying with an exact calculation.

For instance, if we were to fill the lattice with fermions we could go up to to $1/3$ filling (a full flat band) of polarized fermions before a non vanishing inverse-compressibility would set in. It is therefore natural to ask if an exotic, fermionized ground state of bosons could mimic the fermion band filling and thus be favorable to the compressible supersolid state. This can be checked via the calculation of the compressibility as a function of particle doping $\delta \nu$.

Another question raised by the analytical approach above is about the fate of the CDW ordering in the doped state. From the calculation of the density--density correlation function on small clusters we can see if the correlations at the CDW wave vector ${\bf G}_{1}=\frac{2\pi}{a}(1/3,-1/\sqrt{3})$ are indeed enhanced as compared to a uniform state.

\begin{figure}[b]
\includegraphics{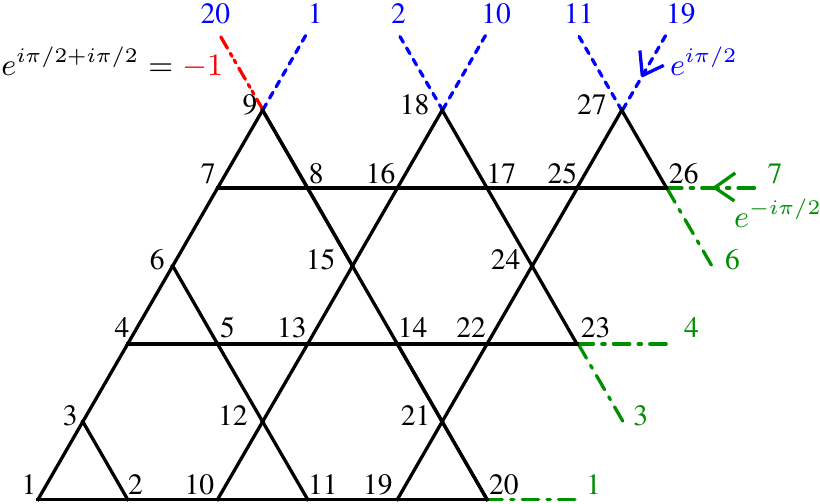}
\caption{
(color online) Boundary conditions for the finite \kg lattice in the presence of a magnetic flux $\Phi_{\pm}\!=\!\pm \pi/2$ through the openings of the torus. The hopping over a blue bond (dashed) is associated with a phase $e^{i\pi/2}$ (the bonds are directed as indicated by the arrow). Particles hopping over green bonds (dashed-dotted) bonds acquire a phase $e^{-i\pi/2}$. The red (dash-dot-dotted) bond connecting site 9 and 20 winds around both circumferences and picks up a phase $-1$.
}
\label{fig:torus}
\end{figure}

We use the ALPS Lanczos application\cite{alps-url,*Albuquerque06} on clusters with a size of 27 sites corresponding to $3\times 3$ unit cells of the \kg lattice. This corresponds to maximal dimension  $d= 807885 $ of the Hilbert space. We perform the calculation with periodic boundary conditions shifted by a phase $\pm\pi/2$ along both $x$ and $y$ directions, which is equivalent to inserting a quarter of a flux quantum through each of the two holes of the torus. The twist serves to eliminate spurious zero energy states which wind around the torus.\cite{Bergman08} The specific gauge choice used to implement these boundary conditions is illustrated in Fig.~\ref{fig:torus} and explained in the caption.

The CDW order on the finite \kg lattice is extracted from the density--density correlation function  $\chi_{ij}=\langle \hat\nu_{i}\hat\nu_{j} \rangle$. Specifically, we define the ``staggered moment":
\begin{equation}
\chi_{\ssm stag}^{2}=\frac{1}{\mathcal{N}}
\sum_{i-j}\chi_{ij}\exp\left[i{\bf G}_{1}\cdot({\bf r}_{i}-{\bf r}_{j})\right],
\end{equation}
The normalization factor $\mathcal{N}$, ensures that the staggered moment is $1$ in the perfect close--packed CDW state illustrated in Fig. \ref{fig:models}.

Fig.~\ref{fig:mf} presents the calculated CDW staggered moment $\chi_{\ssm stag}$ as a function of the doping $\d\nu$ away from the magic filling $\nu_c=1/9$. We find $\chi_{\ssm stag}=1$ at $\nu=\nu_c$ as it should be and the moment falls off at higher filling similarly to the mean-field result. Quantitatively however, the CDW moment from ED is significantly larger  than the mean-field moment. This may seem unusual if we compare for example to the Heisenberg antiferromagnet, where quantum fluctuations beyond the mean-field theory can only reduce the staggered moment. In our case there is the competing superfluid order parameter that facilitates the depletion of the staggered moment even at the mean-field level, however. Quantum fluctuations are generally more effective at suppressing the continuous superfluid order parameter then the discrete CDW. Therefore by depleting the competing superfluid order, quantum fluctuations indirectly act to enhance the CDW order. Indeed we see in Fig.~\ref{fig:mf}, that the superfluid order, measured by the square root of the condensate occupation number, is significantly suppressed in the ED calculation as compared to the mean-field result. Note, that we projected the ED results into the flat band to make connection to the mean-field results. However, the effect of this projection is negligible, as we explicitly show below.
\begin{figure}[t]
\includegraphics{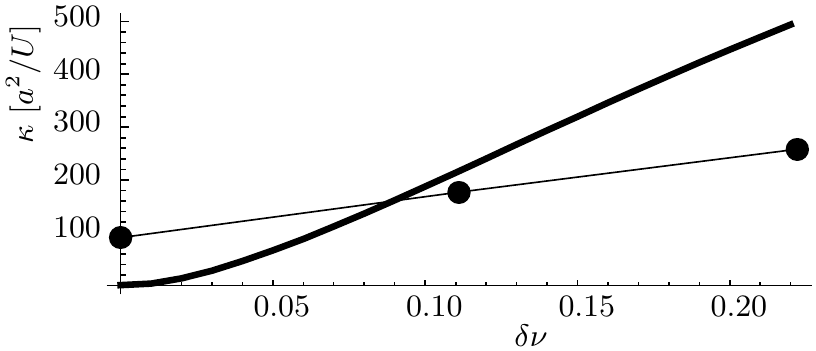}
\caption{
The compressibility as a function of the density $\delta\nu=\nu-\nu_{c}$. The mean-field compressibility is given by the solid line while the dots indicate the results of numerical exact diagonalization. The non-vanishing compressibility of the CDW state at $\nu=\nu_c$ ($\d\nu=0$) is due to finite size of the cluster. Compressibility at $\nu>\nu_c$ is due to the condensate fraction. The fact that the mean-field compressibility is higher than the exact diagonalization result is consistent with the higher condensate density found in the mean-field theory.
}
\label{fig:comp}
\end{figure}

When there are two or more competing order parameters looms the danger of phase separation. To rule out this scenario we shall compute the compressibility $\kappa$ from the ED results using
\begin{equation}
1/\kappa(N) = E_{0}(N+1)-2E_{0}(N)+E_{0}(N-1),
\end{equation}
where $E_{0}(N)$ is the ground state energy of $N$ particles. An instability to phase separation would be signaled by appearance of negative compressibility. From the physics of the flat band below the critical doping $\nu_c=1/9$ we expect the system to be infinitely compressible, or $\kappa^{-1}=0$, in this regime. The CDW state itself at $\nu=1/9$ is incompressible due to the energy gap of order $U$ to addition of one extra-particle on an interstitial site. The danger of phase separation lies when we add a finite density of extra particles to the CDW. However the results shown in Fig.~\ref{fig:comp} imply a state with positive and finite compressibility. Moreover we see a similar trend in both the mean-field and ED results, both showing an increase in  compressibility with increasing density away from the critical density $\nu_c=1/9$. This gives further support to the mean-field picture over both phase separation and the more exotic possibility in which the bosons are fermionized and thereby gain high or infinite compressibility up to filling $\nu=1/3$. The non-zero compressibility at $\delta\nu=0$ is a finite size effect due to the  discreteness in the change of density as we increase particle number on the small lattice.

Finally, we use the ED results to asses the validity of the projection to the lowest band. We calculate the contribution of the kinetic energy to the chemical potential
\begin{equation}
\mu_{\ssm kin}= E_{\ssm kin}(N+1)-E_{\ssm kin}(N),
\end{equation}
where $E_{\ssm kin}(N)$ denotes the expectation value of the kinetic energy in the ground state with $N$ particles. To quantify the overlap with higher bands we take a simple trial state for a single quasi-particle $|\psi\rangle=\cos(\theta)|0\rangle +\sin(\theta)|1\rangle$ where $|0/1\rangle$ denote arbitrary states in the lowest and the first excited band, respectively. With a finite-size band gap $\Delta\approx 0.2\, t$ due to the twisted boundary conditions we can relate the amplitude in the first band to the measured kinetic chemical potential
\begin{equation}
\sin^{2}(\theta) \approx \frac{\mu_{\ssm kin}-2t}{\Delta}.
\end{equation}
The results are given in Tab.~\ref{tab:weights} and we see that for all values of $U$ and $\delta N$ the weight is in the range of a few percent.
\begin{table}[t]
\begin{center}
\begin{ruledtabular}
\begin{tabular}{cccc}
$\delta N$ & $U=0.1\,t$ &$U=0.3\,t$ & $U=0.5\,t$  \\
\hline
1
&
0.2 \%
&
1.6 \%
&
3.7 \%
\\
2
&
0.4 \%
&
3.0 \%
&
6.7 \%
\end{tabular}
\end{ruledtabular}
\end{center}
\caption{
Amplitudes of single particle states in higher bands extracted from the kinetic energy in the exact diagonalization study (see text). The low occupation of the higher bands justifies the effective Hamiltonians derived by projection on the flat lowest band.
}
\label{tab:weights}
\end{table}

\section{Experimental realizations}
\label{sec:experiment}
	
We shall consider two types of systems that can exhibit the physics of bosons in flat bands discussed above. The first class of systems, briefly mentioned in the introduction, are ultracold bosons in optical lattices. The other class includes frustrated antiferromagnets in a large magnetic field. Below we discuss the considerations relevant for the realization of the flat bands in these systems and to the experimental detection of the emergent phases.

\subsection{Cold atoms}

Optical lattice potentials, formed by standing waves of laser light, provide the natural tool to realize interesting Hamiltonians of interacting lattice bosons.\cite{Jaksch98,greiner02} In particular, there have been several suggestions on how to produce a lattice potential with \kg geometry. A direct implementation, proposed by Damski {\em et al.}\cite{Damski05} requires nine pairs of counter-propagating laser beams. Making use of spin dependent lattice potentials it is possible to reduce the number of counter-propagating beams to three.\cite{Ruostekoski09} This method also allows to construct an effective sawtooth geometry.\cite{Huber10b} Very recently optical lattice potentials were produced by an entirely different scheme by Bakr {\em et al.} \cite{Bakr09} In the new scheme the optical potential is projected as a hologram on the ultracold atomic gas, allowing to easily produce any desired lattice geometry.
\begin{figure*}[t]
\begin{center}
\includegraphics{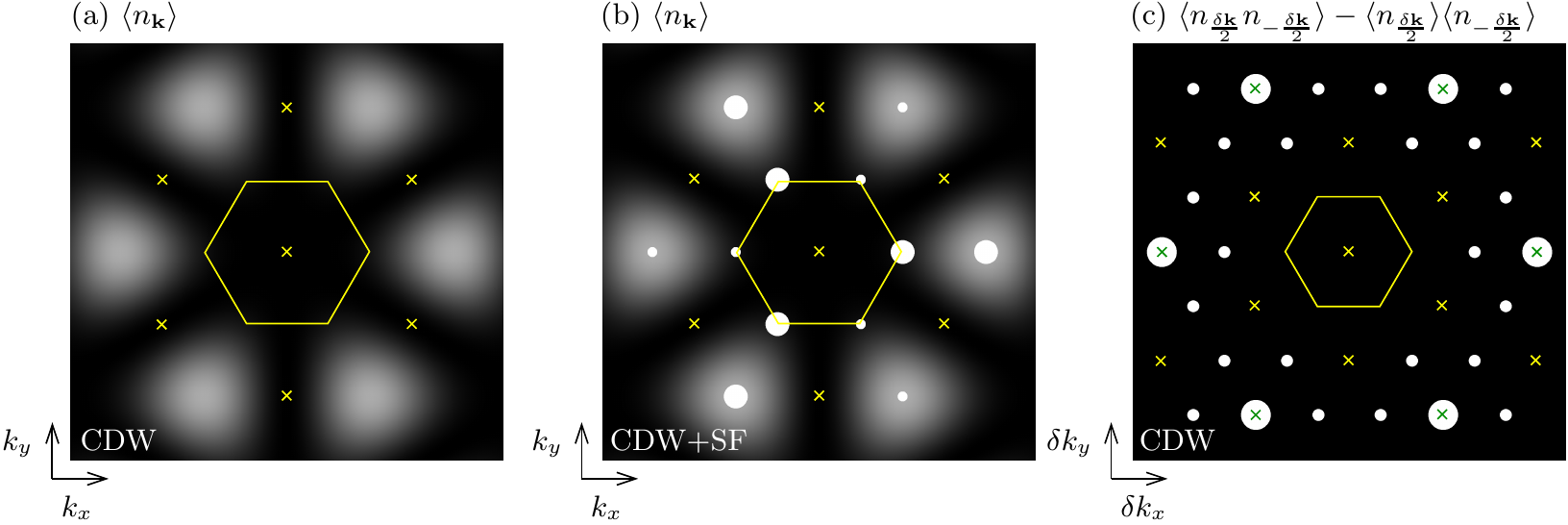}
\end{center}
\caption{
Predictions for time of flight experiments, including momentum distribution and noise correlations for the different phases of the \kg system. (a) The momentum distribution $\av{n_{\bk}}$ in the CDW state at $\nu=\nu_{c}$. The smooth structure stems from the off-diagonal correlations inside hexagons. Destructive interference due to the resonating hexagon wave-function leads to vanishing of $\av{n_\bk}$ on the reciprocal lattice vectors of the original \kg lattice (marked by yellow crosses). The (yellow) hexagon marks the boundaries of the first Brillouin zone.  (b) In the supersolid state at $\nu>\nu_{c}$, sharp delta-function peaks appear in $\av{n_\bk}$ in addition to the  smooth background contribution of the CDW. These condensate peaks appear at the $K$ and $K'$ points and points related to them by the CDW wave-vector. However, points in the reciprocal lattice of the original \kg lattice remain exact zeros of the momentum distribution. In panel (c) we show the noise correlations for the CDW state as a function of the relative momentum $\delta\bk=\bk-\bk'$ for fixed zero average momentum. The delta peaks on the reciprocal vectors of the CDW lattice (peaks away from the yellow crosses that indicate the underlying \kg lattice) indicate the nontrivial ordering in an emergent lattice structure.
}
\label{fig:fq}
\end{figure*}

Another challenging aspect of the experimental implementation is how to obtain frustrated hopping of the bosons on the lattice. Usual quantum mechanical tunneling of bosons between the wells naturally gives rise to a negative hopping matrix element, which does not lead to frustration. However, several clever schemes have been proposed to invert the hopping matrix element and thus produce the desired frustration. One scheme proposed by Eckardt {\em et. al.}\cite{Eckardt10} achieves this by introducing a fast time-periodic lattice acceleration. With proper design of the time dependent acceleration, the resulting Floquet Hamiltonian has an inverted sign of the hopping. The general idea of tuning of the effective hopping with dynamical modulation\cite{Dunlap86,Eckardt05} was successfully demonstrated in experiment.\cite{Lignier07}

A second scheme to invert the hopping is tailored to implementations of the lattice with spin dependent optical potentials. The phase of the tunneling matrix elements can then be tuned by controlling the phase of a laser field that induces transitions between the two relevant spin states. \cite{Huber10b}

A third and completely different route for achieving the frustrated hopping relies on a mapping between the low-energy dynamics of a target Hamiltonian $H$ and the  dynamics in the highest energy manifold of $-H$.\cite{Sorensen10} In our case $H$ is the desired Hamiltonian, which describes repulsively interacting bosons with frustrated hopping and we are interested in the low-energy dynamics on the lowest band (i.e. the flat band). While this is hard to produce, it is rather straightforward to realize $-H$, the un-frustrated model with attractive interactions. If this system can be initialized with all particles in the flat upper band, then the ensuing dynamics will be identical to the low-energy dynamics of the frustrated model $H$. Decay of particles from the flat band to the lowest band is exponentially suppressed in the large parameter $\hbar\omega_{0}/t$ where $\hbar\omega_{0}$ is the gap to higher Bloch bands.

Given an experimental realization of the model (\ref{eqn:bosonic-hamiltonian}) of frustrated bosons on the \kg or \st lattice, the ground state and low-energy properties can be measured with standard probes. Here we concentrate on the time of flight expansion image which gives information on the ground state correlations. We note that the excitation spectrum can be studied using Bragg-spectroscopy\cite{Stenger99} or lattice modulation spectroscopy.\cite{Stoferle04,Huber07}

The time of flight image lends direct access to the ground state momentum distribution $\av{n_{\bf k}}=\langle b_{\bf k}^{\dag} b_{\bf k}\rangle$ and the noise correlations in the image reflect the momentum space correlations: $\mathcal{G}(\bk,\bk')=\av{n_{\bf k}n_{\bf k'}}-\av{n_{\bf k}}\av{n_{\bf k'}}$.\cite{Altman04b} As we show below, these two observables complement each other to give detailed information on the emergent ground state of the frustrated bosons.

We start by considering the momentum distribution in the pure CDW state on the \kg lattice at $\nu=1/9$. The delocalization of bosons inside hexagons gives rise to distinct structure in $\av{n_\bk}$, albeit no sharp delta-function features in absence of long-range superfluid order. To obtain the structure we rewrite the momentum distribution in a real space representation
\bea
\av{n_\bk}={1\over M}\sum_{ij}e^{i\bk\cdot (\bx_i-\bx_j)}\av{b\yd_i b\nd_j}.
\label{nk}
\eea
where $M$ is the number of \kg lattice points.  Now using the facts that $b_j$ annihilates the CDW state unless $j$ belongs to one of the occupied hexagons and that the off-diagonal correlations are non-zero only within a single hexagon we can rewrite (\ref{nk}) as
\bea
\av{n_\bk}_{\ssm CDW}&=&
{1\over M}
\sum_{\bR_{i}}\sum_{\alpha\alpha'}e^{i\bk\cdot(\br_\a-\br_{\a'})} \av{b\yd_\a b\nd_{\a'}}
\nn\\
&=&{1\over 54}\sum_{\alpha\alpha'}e^{i\bk\cdot(\br_\a-\br_{\a'})}(-1)^{\a+\a'}
\eea
where $\bR_{i}$ are the centers of the occupied hexagons, $\a$ labels the site of a hexagon and the vectors $\br_\a$ give the locations of the hexagon points relative to its center. Plugging in these vectors finally gives the simple expression
\be
\nn
\av{n_\bk}_{\ssm CDW}=
\frac{8}{27}\sin^{2}\bigg(\frac{k_{x}}{4}\bigg)
\left[
\cos\bigg(\frac{k_{x}}{4}\bigg)-\cos\bigg(\frac{\sqrt{3}k_{x}}{4}\bigg)
\right]^{2},
\ee
which is plotted in Fig.~\ref{fig:fq}(a). It is interesting to note that this function has zeros on the reciprocal lattice vectors of the \kg lattice (more precisely those of the underlying triangular Bravais lattice). The maxima of the function are located on the reciprocal lattice of the emergent CDW structure, but not on those that also belong to the reciprocal of the original lattice where $\av{n_\bk}=0$. Note that the calculated momentum distribution does not decay at large wave-vectors $k$. This is because we took  lattice wave-functions to be strictly confined to lattice points. In reality the momentum distribution will have a decaying envelope, reflecting they are composed of local wave functions with finite extent

To obtain the approximate momentum distribution in the supersolid phase at filling $\nu>1/9$ we apply the prescription (\ref{nk}) to the mean-field wave-function
\be
\ket{\Psi(\{\t_i,\f_i\})}=\prod_j\left(\sin(\t_j/2)+ e^{i\f_j}\cos(\t_j/2) {\rm W}\yd_j\right)\ket{0}.
\nn
\ee
Here $j$ runs over unit cells, or all hexagons in the \kg lattice and ${\rm W}\yd_j$ creates the resonating state localized on hexagon $j$ (\ref{eqn:hexagon}). As shown in section \ref{sec:kgmean}, $\t_j$ and $\f_j$ form a three-unit cell periodic structure which amounts to a superfluid order parameter at the $K$ and $K'$ points of the original Brillouin zone in addition to the CDW order. Because of the off-diagonal long-range order, delta function peaks appear in $\av{n_\bk}$  on top of the smooth structure caused by the hexagon CDW.  Figure~\ref{fig:fq}(b) shows the condensate peaks which appear at the $K,K'$ points of the \kg Brillouin zone and at other points connected to them by the wave-vectors of the CDW. However, the form factor controlling the relative strength of the different peaks is non trivial and most interestingly implies that some peaks have zero weight. Specifically, the reciprocal lattice vectors of the \kg lattice remain zeros of $\av{n_\bk}$ although they are also connected to the $K$ and $K'$ points by CDW wave-vectors. The zeros are the result of destructive interference from the staggered phases of the resonating hexagon states.

The noise correlations in the CDW state are computed, similarly to the momentum distribution, by expressing them in terms of the real space lattice operators (see Ref.~\onlinecite{Altman04b})
\bea
\mathcal{G}(\bk,\bk')&=&\sum_{iji'j'}e^{i\bk\cdot(\bx_i-\bx_j)+i\bk'\cdot(\bx_{i'}-\bx_{j'})}\nn\\
&&\times\left(\d_{ji'}\av{b\yd_i b\nd_{j'}}+\av{b\yd_i b\nd_{j'}}\av{b\yd_{i'} b\nd_{j}}\right).
\eea
The fact that $\av{b\yd_i b\nd_j}$ is non zero, in the pure CDW state, only if $i$ and $j$ belong to the same {\em occupied} hexagon, allows us to write $\mathcal{G}(\bk,\bk')$ in terms of the sites of a single hexagon
\bea
\mathcal{G}(\bk,\bk')&=&\d_{\d \bk,\bQ} f(\bK,\d \bk)+\d_{\d \bk,\bG}f^{2}(\bK,\d \bk)\nn\\
f(\bK,\d \bk)&\equiv&{1\over 6}\sum_{\a,\a'=1}^6(-1)^{\a+\a'}e^{i\bK\cdot(\br_\a-\br_{\a'})+i\d \bk\cdot {\br_\a+\br_{\a'}\over 2}}
\nn
\eea
where $\d \bk\equiv \bk-\bk'$ and $\bK\equiv (\bk+\bk')/2$, $\bQ$ denote the reciprocal lattice vectors of the original \kg lattice, and $\bG$ are the wave-vectors of the CDW order.

Note the sharp peaks in the noise correlations, which appear when $\bk-\bk'$ is equal to a reciprocal lattice vector of the emergent CDW structure. This is different from a Mott insulating state, where lattice symmetry is not broken and therefore peaks are observed when $\bk-\bk'$ is equal to a reciprocal lattice vector of the optical lattice potential.\cite{Foelling05} The noise correlations thus provide a clear signature of the CDW order.

The form factor which controls the relative strength of the peaks at different reciprocal lattice vectors, depends on both the relative momentum $\d\bk$ and the average momentum $\bK$. The dependence on the average momentum is determined by the inter-site coherence within a unit cell of the CDW, information that is already available in $\av{n_\bk}$. It is therefore useful to fix $\bK$, as done in Fig. \ref{fig:fq}(c), in order to isolate the dependence of the form factor on $\d\bk$ and thereby obtain information on the density profile within a unit cell.

\subsection{Quantum magnets}

Due to the analogy between bosons and quantum spins, the question of Bose condensation in flat bands has a lot in common with the problem of frustrated quantum magnets.\cite{Moessner01b,Wannier50,Houtapple50,Syozi51,Villain80a,Chalker92,Ramirez94,Moessner98,Gardner10} The frustrated geometry gives rise to an extensive degeneracy that inhibits formation of broken spin symmetry, just as the flat bands inhibit Bose condensation.

In fact, there is a direct mapping between antiferromagnets, which are nearly fully polarized by an external magnetic field, and the model (\ref{eqn:bosonic-hamiltonian}) of weakly interacting lattice bosons. A spin flip in the otherwise polarized state behaves as a bosonic excitation, known as a magnon.\cite{Bloch30,Dyson56} The magnon hops between nearest neighbor sites with a {\em positive} matrix element due to the antiferromagnetic exchange, whereas single ion anisotropy terms $\propto (S^z_i)^2$ in magnets with spin $S\ge 1$ translate to on site interactions between magnons.

Decreasing the magnetic field is like tuning the chemical potential of the magnons and can lead to Bose condensation.\cite{Rice02,Giamarchi99,Giamarchi08} At very high magnetic fields compared to the Curie Weiss temperature of the antiferromagnet, the spins are fully polarized and magnons are gapped. However, below a critical field the gap closes and the magnons are expected to condense giving rise to long-range spin correlations in the spin axes perpendicular to the magnetic field. Such a transition has been observed recently in the spin ladder system (Hpip)$_{2}$CuBr$_{4}$.\cite{Thielemann09} In the spin ladder the magnon band has a unique minimum energy. The approach developed in this paper becomes useful when the lattice geometry is sufficiently frustrated to induce a flat magnon dispersion,\cite{Zhitomirsky05} which inhibits straightforward magnon condensation.

An interesting system from this viewpoint is the organic material $m$-MPYNN$\cdot$BF$_{4}$,\cite{Awaga94,Wada97} which forms a \kg structure of effective spin-1 degrees of freedom. Fortunately it also has a low Curie-Weiss temperature ($\Theta_{\ssm CW}\!\approx\!3\,{\rm K}$), which allows to completely polarize the spins with reasonable magnetic fields. Indeed, NMR spin relaxation measurements indicate the presence of a spin gap above a field $h_{c2}\sim 4\,{\rm T}$.\cite{Fujii02,*Fujii03} The close packed hexagon lattice should show up as a magnetization plateau at $8/9$ of full polarization at magnetic fields slightly below the saturation field. Our results pertain to the nature of the ground state with magnetization below $8/9$ per spin.

The detailed ordering structure can in principle be obtained from elastic neutron scattering experiments which measure the spin correlation function. In particular, the $x-y$ component of the static structure factor $\av{S^+_q S^-_q}$ can be mapped directly to the momentum distribution of the bosons discussed above and shown in Fig. \ref{fig:fq}(b) (see App.~\ref{app:spin-boson} for the details of the spin-boson mapping). The $z$-component of the spin structure factor can be directly mapped to the Fourier transform of the density correlations. It will show peaks at the CDW wave-vectors modulated by a form factor which depends on the density profile within a unit cell. In that sense neutron scattering will give information similar to the noise correlations in time of light images of ultracold atoms at a fixed value of the center of mass momentum (See Fig. \ref{fig:fq}).

Finally, we note that in order to apply the results of our study to $m$-MPYNN$\cdot$BF$_{4}$, it may be necessary to take into account the structural distortion that occurs below  $T_{s}\approx 130\, {\rm K}$. This distortion leads to anisotropy of the magnetic exchange along different directions of the \kg lattice.\cite{Kambe04,*Kambe04a} As long as the anisotropy is weak compared to the effective magnon interactions then the use of the effective Hamiltonian restricted to the lowest band is still justified, albeit with spatially anisotropic interactions.

\section{Conclusions and outlook}

We have developed a general scheme to derive low-energy effective Hamiltonians for bosons on the flat Bloch bands of frustrated lattices. The essential idea is to project the Hamiltonian on the flat band by formulating the interaction term using a complete basis of localized Wannier states of the flat band. The resulting model contains both diagonal interaction terms and off-diagonal (assisted) hopping terms, all proportional to the interaction parameter in the original model. Most importantly, the effective model is in most cases no longer highly frustrated and therefore amenable to analysis by mean-field theory and other standard theoretical methods. Such analysis is directly useful in the description of ultracold atoms in optical lattices with highly frustrated hopping and of frustrated quantum magnets in high magnetic fields, where magnons behave as dilute bosons.

Using the projection scheme, we investigated the Bose Hubbard model on the frustrated \st and \kg lattices. In both cases it is possible to construct exact many-body ground states composed of nonoverlapping configurations of localized states, which lie entirely in the flat band\cite{Schulenburg02,Zhitomirsky04} (see Fig. \ref{fig:strictly}). The existence of a large number of such configurations prevents any ordering at zero temperature. Above the critical lattice filling $\nu_c$, at which the localized states form a closely packed density wave, the exact construction fails and our approach is useful in identifying the emergent ground state and low-energy excitations.

In the one-dimensional\st lattice we show that the CDW is immediately destroyed by proliferation of domain walls upon increasing the density from the magic filling $\nu_c=1/4$. Universal properties, e.g., the power-law peaks in the static structure factor, at filling $\nu=1/4+\d\nu$ are described by the long-wavelength theory of the commensurate-incommensurate transition.

In the \kg lattice, the CDW remains stable upon increasing the density from $\nu_c=1/9$.  The added bosons hop between interstitial sites of the CDW and form a condensate at the $K$ and $K'$ points of the Brillouin zone of the \kg lattice.  We considered the signatures of this phase in a realization with ultracold atoms in optical lattices. The predicted time of flight image is shown in Fig. \ref{fig:fq}(b) shows the sharp peaks indicating the presence of a condensate at the right wave-vector. Most strikingly however, as a result of destructive interference from the localized hexagon wave-functions, we find zeros of the momentum-distribution on the reciprocal lattice vectors of the original \kg lattice.

A general feature of Bose condensates in flat bands of frustrated lattices is that condensation is entirely interaction driven. It is a result of the emergent dynamics induced inside the flat band by the weak interactions, which is directly captured in the projection scheme. In particular both the condensation energy and superfluid stiffness are proportional to the Hubbard interaction $U$ and not to the kinetic energy as in usual condensates. In a similar way, due to effective interactions generated in the low-energy model, charge density wave states and supersolids are stabilized even in absence of long-range interactions in the microscopic model.

It would be interesting to apply the ideas developed in this paper to frustrated antiferromagnets such as the organic salt $m$-MPYNN$\cdot$BF$_{4}$. Such analysis should include coupling to the anisotropic lattice distortion. Furthermore, nearest neighbor magnon interactions, which are naturally generated by the antiferromagnetic magnetic exchange, are expected to give rise to more magnetization plateaux. An intriguing question then is how the supersolid phase we find connects with the plateaux at lower magnetization? Using the projection method to investigate the compressible phases, which connect between lower plateaux, and perhaps finally with the spin gapped phase at zero field, offers a new viewpoint on the problem of frustrated quantum magnets.

\acknowledgments

We acknowledge stimulating discussions with John Chalker, Erez Berg, Doron Bergmann, Immanuel Bloch, Andrew Daley, Thierry Giamarchi, David Huse, Andreas R\"uegg, Ronny Thomale, Matthias Troyer, and Peter Zoller. We thank T. Matsushita and K. Awaga for sharing their unpublished high-field data with us. This work was supported in part by ISF, US-Israel BSF, and the Minerva foundation. S.D.H. acknowledges financial support by the Center for Theoretical Studies at ETH Z\"urich and the Swiss Society of Friends of the Weizmann Institute of Science.

\appendix

\section{Details of the projection to the flat band}
\label{app:projection}

\begin{figure}[t]
\includegraphics{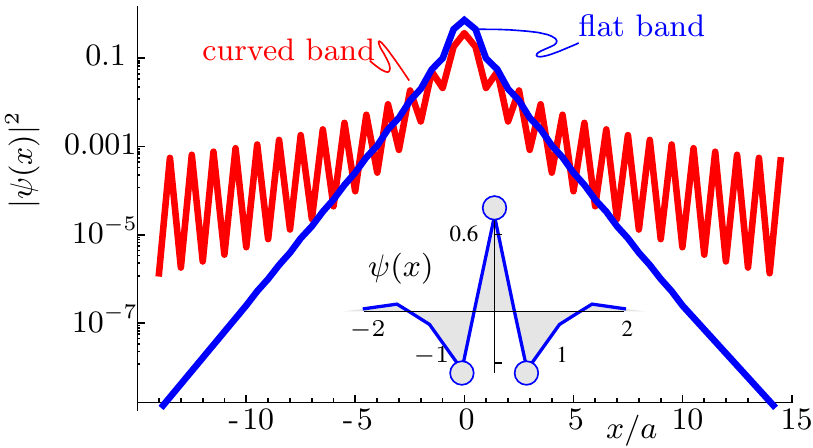}
\caption{
(color online) Wannier functions (probability distribution) as a function of position on the \st lattice with $t'/t=\sqrt{2}$. The excited band has an oscillatory slowly decaying tail, while the flat-band Wannier functions decay with a localization length $\xi\approx \log(2.15) a$. The inset shows that locally, the Wannier function of the flat band is almost identical to the strictly localized V states indicated by the filled circles.
}
\label{fig:wannier-st}
\end{figure}

The explicit expressions for the Wannier functions of the flat band of the \st lattice can be written in terms of complete elliptic integrals of the first and second kind, $K(m)$ and $E(m)$, respectively.

Specifically, the amplitudes used in the derivation of the effective Hamiltonian (\ref{eqn:stspinham}) are given by
\begin{align}
w_{B}(0) &= -\frac{2}{3\sqrt{\pi}} K(2/3), \\
w_{B}(a) &= -\frac{2}{3\sqrt{\pi}}(3E(2/3)+2K(2/3)),\\
w_{A}(0) & = \sqrt{\frac{2}{3}}\frac{1}{\pi}(3E(2/3)-K(2/3)),
\end{align}
\begin{align}
w_{A}(a) & = \sqrt{\frac{2}{3}}\frac{1}{\pi}(3K(2/3)-5E(2/3)).
\end{align}

The resulting interaction constants are of the form:
\begin{align}
I^{z}/U&\approx
8 w_{B}^{2}(0)w_{B}^{2}(a)+
8 w_{A}^{2}(0)w_{A}^{2}(a)+
4 w_{A}^{4}(0)
\nonumber
\\&\approx \phantom{-}0.112, \\
I^{\ssm dw}/U&\approx
4 w_{B}^{3}(a)w_{B}(0)+
4 w_{A}^{3}(0)w_{A}(a)+
4 w_{A}^{3}(a)w_{A}(0)
\nonumber
\\&  \approx  -0.025, \\
I^{\ssm sp}/U&\approx
4 w_{B}^{2}(a)w_{B}^{2}(0)+
8 w_{A}^{3}(0)w_{A}(a)
\nonumber
\\&\approx  -0.011,\\
\tilde U/U &\approx
 w_{B}^4(0) + 2 w_{B}(a)^4 + 2 w_{A}^4(0) + 2 w_{A}^4(a)
\nonumber
\\& \approx \phantom{-}0.40.
\end{align}

Fig. \ref{fig:wannier-st} shows the spatial dependence of the Wannier functions of both the lower flat band and the upper band. Note the strong localization of the Wannier function of the flat band. A zoom-in on the short range structure of this function on-site and on nearest neighbors shows that it is shows that it is almost identical to the strictly localized V states of the \st lattice.

\begin{figure}[bt]
\includegraphics{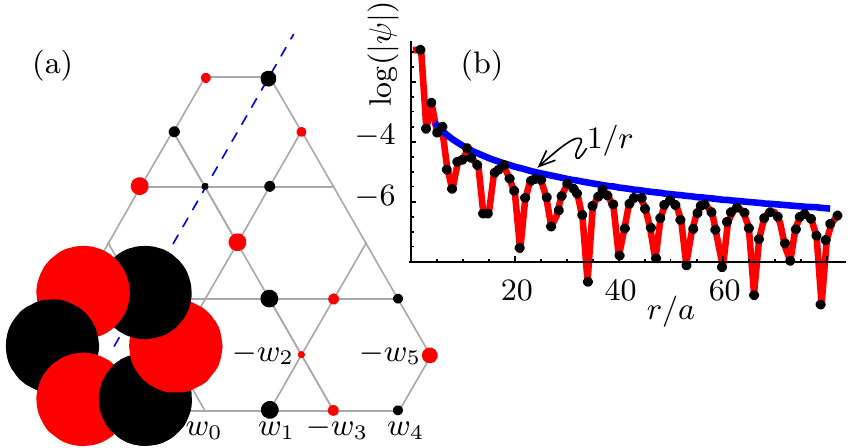}
\caption{
(color online) Wannier function for the flat band in the \kg lattice. (a) Wave-function amplitudes shown on the two-dimensionallattice, where the size of the dots represents the weight and the color (shading) the denotes the sign $+$ or $-$ of the wave-function. The definitions of the weights $w_{1}\dots w_{5}$ are indicated. (b) Algebraic decay of the Wannier function as $\sim 1/|r|$ along the high symmetry direction marked by the dashed line in (a).
}
\label{fig:wannier-kg}
\end{figure}

 Fig.~\ref{fig:wannier-kg} shows the spatial behavior of the flat band Wannier functions of the \kg lattice as defined in (\ref{eqn:kgwannier}). Due to the touching band, they do not fall of exponentially but decay algebraically, cf. Fig.~\ref{fig:wannier-kg}(b).\cite{Kohn59,Des-Cloizeaux63,Des-Cloizeaux64a,Des-Cloizeaux64}

Coefficients in Hamiltonian (\ref{eqn:kgeff}) result from the overlap of Wannier functions at different sites. The explicit expressions for the first few terms are given by
\begin{align}
\tilde U/U &\approx 6w_{0}^{4}\approx 0.144, \\
I^{z}/U & \approx w_{0}^4 + 4 w_{0}^2 w_{1}^2 + 2 w_{0}^2 w_{2}^2 \approx 0.028, \\
I^{xy}_{1}/U &\approx 2 w_{0}^3 w_{1} \approx 0.009, \quad
I^{xy}_{2} = I^{xy}_{1}/2,\\
I^{xy}_{3}/U &\approx 2 w_{0}^3 w_{2} \approx 0.003, \quad
I^{xy}_{4} = I^{xy}_{3}/2
 \end{align}
where the amplitudes $w_\a$ are defined in Fig. \ref{fig:wannier-kg}.

\section{Spin--boson mapping}
\label{app:spin-boson}

In this appendix we describe details of the mapping from a spin-$S$ system to a bosonic Hamiltonian. We consider the following general spin Hamiltonian
\begin{equation}
\label{eqn:spin1-hamiltonian}
H=J\sum_{\langle ij\rangle}
\bigl[
S_{i}^{+}S_{j}^{-}+S_{i}^{-}S_{j}^{+}+\lambda S_{i}^{z}S_{j}^{z}
\bigr]\!+\!\sum_{i}\bigl[D(S_{i}^{z})^{2} - h S_{i}^{z} \bigr].
\end{equation}
 which includes anisotropic magnetic exchange interactions as well as a single ion anisotropy $D$. The latter is relevant for $S>1/2$ where two magnons can reside on the same site; we concentrate on this regime in the following. This Hamiltonian has been studied extensively at small magnetic field $h$ and $S=1$.\cite{Wen03,Damle06,Xu05a,Xu07,Levin07}

Here, we focus on the large field limit. The fact that spins are nearly polarized in the $z$ direction suggests to construct a bosonic representation of the small fluctuations (magnons) about the fully polarized state using the Holstein--Primakoff expansion.\cite{Auerbach94} To quadratic order in the magnon occupation the mapping between the spins and bosons is:
\begin{align}
\nn
S^{+}_{i} &\approx \sqrt{2S}\left(1-\frac{b_{i}^{\dag}b_{i}^{\pdag}}{4S}\right) b_{i},
\\
\label{eqn:hpk}
S^{-}_{i} &\approx b_{i}^{\dag}\sqrt{2S}\left(1-\frac{b_{i}^{\dag}b_{i}^{\pdag}}{4S}\right),\\
\nn
S^{z}_{i} &= (S-b_{i}^{\dag}b_{i}^{\pdag}).
\end{align}
 Expressing Hamiltonian (\ref{eqn:spin1-hamiltonian}) using the Holstein-Primakoff bosons and keeping terms to leading order in $1/S$ we finally obtain the bosonic Hamiltonian
\begin{align}
\label{eqn:magnon}
H&=JS \sum_{\langle ij\rangle}
\bigl[ b_{i}^{\dag}b_{j}^{\pdag} + {\rm H.c}\bigr]
+\sum_{i}\mu_{\ssm eff} \hat\nu_{i}+D\sum_{i}\hat\nu_{i}^{2}
\\
&\phantom{=}-
\nn
J\sum_{\langle ij\rangle}
\left[\frac{1}{2}\left(
\hat \nu_{i}b_{i}^{\dag}b_{j}+{\rm H.c}
\right)+\lambda \hat \nu_{i}\hat\nu_{j}
\right] + \mathcal{O}(1/S^{2}).
\end{align}
Here, $\hat \nu_{i}=b_{i}^{\dag}b_{i}^{\pdag}$ denotes the density operator. The effective chemical potential for the magnons here is given by $\mu_{\ssm eff}=SD+2z\lambda JS+h$. The Bose-Hubbard model (\ref{eqn:bosonic-hamiltonian}), which was considered in this paper, is only the first line of (\ref{eqn:magnon}). The second line includes additional nearest neighbor interactions and assisted hopping terms which stem from the antiferromagnetic exchange interactions. Note however, that both terms in the second line are vanishing in the exact CDW state and can be projected into the flat band along the same lines as the simple on-site interaction $\propto D$.

\end{document}